\documentclass[aps,pra,twocolumn,amsmath,amssymb,showpacs,floatfix]{revtex4}
\usepackage{graphicx}
\begin{document}


\title{Exact Performance of Concatenated Quantum Codes}

\author{Benjamin Rahn} 
\email{brahn@caltech.edu} 
\author{Andrew C. Doherty}
\author{Hideo Mabuchi}
\affiliation{Institute for Quantum
Information, California Institute of Technology}

\date{June 10, 2002}

\begin{abstract}
When a logical qubit is protected using a quantum error-correcting code,
the net effect of coding, decoherence (a physical channel acting on
qubits in the codeword) and recovery can be represented exactly by an
effective channel acting directly on the logical qubit.  In this paper
we describe a procedure for deriving the map between physical and
effective channels that results from a given coding and recovery
procedure.  We show that the map for a concatenation of codes is given
by the composition of the maps for the constituent codes.  This perspective
leads to an efficient means for calculating the exact performance of
quantum codes with arbitrary levels of concatenation.  We present
explicit results for single-bit Pauli channels.  For certain codes under
the symmetric depolarizing channel, we use the coding maps to compute
exact threshold error probabilities for achievability of perfect
fidelity in the infinite concatenation limit.
\end{abstract}


\maketitle

\def\expval[#1]{\langle #1 \rangle}
\def\tr{\mathrm{tr}}
\def\comment{\bf \footnotesize}
\def\ket[#1]{\vert #1 \rangle}
\def\bra[#1]{\langle #1 \vert}
\def\braket[#1,#2]{\langle #1 \vert #2 \rangle}
\def\ketbra[#1,#2]{\vert #1 \rangle \langle #2 \vert}
\def\Nop{\mathcal{N}}
\def\Gop{\mathcal{G}}
\def\Eop{\mathcal{E}}
\def\Dop{\mathcal{D}}
\def\Ndep{\mathcal{N}^\mathrm{dep}}
\setlength{\parskip}{0pt}
\setlength{\textfloatsep}{6pt}
\setlength{\floatsep}{7pt}
\def\0bar{\overline{0}}
\def\1bar{\overline{1}}
\def\hide[#1]{}
\def\inn{\mathrm{in}}
\def\out{\mathrm{out}}
\def\Shor{\mathrm{Shor}}
\def\btf{\mathrm{bf}}
\def\phf{\mathrm{pf}}
\def\Steane{\mathrm{Steane}}
\def\Five{\mathrm{Five}}
\def\wt{\widetilde}

\section{Introduction}

The methods of quantum error correction
\cite{qecc_original,steane,qecc_background} have, in principle, provided a
means for suppressing destructive decoherence in quantum computer
memories and quantum communication channels.  In practice, however, a
finite-sized error-correcting code can only protect against a subset of
possible errors; one expects that protected information will still
degrade, albeit to a lesser degree.  The problem of characterizing a
quantum code's performance could thus be phrased as follows: what are
the effective noise dynamics of the encoded information that result from
the physical noise dynamics in the computing or communication device?

One could address this question by direct simulation of the quantum
dynamics and coding procedure.  However, for codes of non-trivial size,
this approach rapidly becomes intractable.  For example, in studies of
fault-tolerance \cite{ftc_background} one often considers families of
concatenated codes \cite{concatenation_original,qecc_background}.  An
$N$-qubit code concatenated with itself $\ell$ times yields an
$N^\ell$-qubit code, providing better error resistance with increasing
$\ell$.  For even modest values of $N$ and $\ell$, simulation of the
resulting $2^{(N^\ell)}$-dimensional Hilbert space requires massive
computational resources; using simulation to find the asymptotic
performance as $\ell \rightarrow \infty$ (as required for fault-tolerant
applications) is simply not on option.

Instead, a quantum code is often characterized by the set of discrete
errors that it can perfectly correct \cite{theory_qecc}.  For example,
the Shor nine-bit code \cite{qecc_original} was designed to perfectly
correct arbitrary decoherence acting on a single bit in the nine-bit
register.  Typical analyses of such codes implicitly assume that the
physical dynamics can be described by single-bit errors occurring at some
probabilistic rate; if this rate is small (e.g. $O(p)$ for $p\ll 1$),
the probability that these errors will accumulate into a multi-bit
uncorrectable error is also small (e.g. $O(p^2)$).  This type of
leading-order analysis is limited to a weak-noise regime, and to error
models strongly resembling the errors against which the code protects.
Outside of this regime, these approximation methods fail to accurately
describe the evolution of the encoded information.  

In this work we take a different approach to characterizing
error-correcting codes, which leads to a simple, exact analysis for
arbitrary error models.  As suggested above, a code transforms the
physical dynamics of the device into the effective dynamics of the
encoded information.  In section \ref{sec:effective} we derive this
transformation for arbitrary noise, and present a compact method for
its calculation.

In the case of identical, uncorrelated noise on individual qubits, this
notion becomes particularly natural: encoding a logical qubit in several
physical qubits yields an evolution less noisy than if the logical qubit
had been stored, unencoded, in a single physical qubit.  Thus a code
acts a map on the space of qubit dynamics, mapping the dynamics of a
single \textit{physical} qubit to the dynamics of the encoded
\textit{logical} qubit.  In section \ref{sec:coding_map} we show how to
calculate this map, and in section \ref{sec:concatenated} we use these
maps to dramatically simplify the calculation of effective dynamics for
concatenated codes when the physical dynamics do not couple code blocks.

In section \ref{sec:diagonal}, we restrict our attention to uncorrelated
single-bit Pauli errors, and in section \ref{sec:performance} we
calculate the exact performance of several codes of interest under these
error models.  Finally, in section \ref{sec:thresholds} we use the
coding maps to calculate the performance of certain concatenated codes,
and find the exact threshold error probability for perfect fidelity in
the infinite concatenation limit.  These thresholds serve as important
figures of merit for concatenation schemes, and for the codes considered
here we find that the traditional approximate methods underestimate
these thresholds by up to $44\%$.  Section \ref{sec:conclusion}
concludes, suggesting potential future applications for these
techniques.

\section{Describing Code Performance with Effective 
Channels\label{sec:effective}}

In this section we first describe error-correcting codes using a
language that will facilitate the subsequent development.  We will then
present our method for exactly describing the effective dynamics of the
encoded information.  Though for clarity we restrict our discussion
to codes storing a single qubit (sometimes called $k=1$ codes) all of
the presented methods generalize naturally to codes storing quantum
information of arbitrary dimension.

As an important preliminary, it can be argued \cite{qecc_background,
kraus} that all physically possible transformations taking quantum
states $\rho$ on a Hilbert space $\mathcal{H}$ to states $\rho'$ on
a Hilbert space $\mathcal{H}'$ may be written in the following
form:
\begin{equation}
\rho \rightarrow \rho' = \sum_j A_j \rho A_j^\dagger
\;\;\;\mathrm{with}\;\;\;
\sum_j A_j^\dagger A_j = \mathbf{1}
\end{equation}
where the $A_j$ are linear operators from $\mathcal{H}$ to
$\mathcal{H}'$ and $\mathbf{1}$ denotes the identity operator on
$\mathcal{H}$.  Such transformations are called \textit{quantum
operations} or \textit{channels}, and are necessarily linear,
trace-preserving, and completely positive.  It is easy to see that the
composition of quantum operations is also a quantum operation.  (One
also sees definitions requiring only $\sum_j A_j^\dagger A_j \leq
\mathbf{1}$, corresponding to the weaker requirement that a quantum
operation be trace non-increasing rather than trace-preserving.
However, the requirement of trace-preservation is better suited to our
purposes here.  See \cite{qecc_background} for a discussion of the
distinction.)

\subsection{The Error-Correction Process\label{sec:qec_basics}}

\begin{figure}
\includegraphics{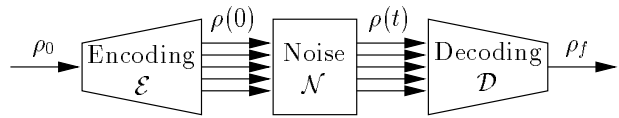}
\caption{The error-correction process: a logical single-qubit state
$\rho_0$ is encoded in an $N$-qubit register as $\rho(0)$.  A noise
process transforms the register to state $\rho(t)$, which is then
decoded to yield the logical single-qubit state $\rho_f$.}
\label{fig:qec_process}
\end{figure}

The error-correction process, consisting of encoding, noise, and
decoding, is depicted in Figure \ref{fig:qec_process}; we consider each
stage in turn.  An $N$-qubit code $C$ uses a register of $N$ qubits to
encode a single logical qubit $\alpha\ket[0]+\beta\ket[1]$ by preparing
the register in the state $\alpha\ket[\0bar]+\beta\ket[\1bar]$, where
$\ket[\0bar]$ and $\ket[\1bar]$ are orthogonal states in the
$2^N$-dimensional Hilbert space of the register.  The codespace (i.e.,
the space of initial register states) is spanned by these two states.
In what follows it will be convenient to describe states by density
matrices: let the logical qubit be given by $\rho_0$ and the initial
register state by $\rho(0)$.  Writing
$B=\ketbra[\0bar,0]+\ketbra[\1bar,1]$, the \textit{encoding operation}
$\Eop: \rho_0 \rightarrow \rho(0)$ is given by
\begin{equation}
\label{eqn:encoding_kraus}
\rho(0) = \Eop[\rho_0] = B\rho_0 B^\dagger.
\end{equation}
As $B^\dagger B = \ketbra[0,0] + \ketbra[1,1] = \mathbf{1}$, $\Eop$ is a
quantum operation.

After the encoding, the register state evolves due to some noise
dynamics.  In the setting of a quantum computer memory, the dynamics are
continuous in time; assuming evolution for a time $t$, we have $\rho(t)
= \Nop_t[\rho(0)]$ with $\Nop_t$ a quantum operation depending continuously
on $t$.  (For master equation evolution $\dot{\rho} =
\mathcal{L}[\rho]$, we have $\Nop_t = e^{\mathcal{L}t}$.)  We will often
omit the subscript $t$ and simply write $\Nop$.  In the setting of a
quantum communication channel, the noise process is usually given by the
discrete application of a quantum operation $\Nop$; thus if the transmitted
state is $\rho(0)$, the received state is $\Nop[\rho(0)]$, which we
write as $\rho(t)$ for consistency.

After the noise process, an attempt is made to recover the initial
register state $\rho(0)$ from the current register state $\rho(t)$ by
applying a quantum operation $\mathcal{R}$, which may be written as
\begin{equation}
\label{eqn:recovery_kraus}
\mathcal{R}[\rho(t)] = \sum_j A_j \rho(t) A_j^\dagger
\;\;\;\mathrm{with}\;\;\;
\sum_j A_j^\dagger A_j = \mathbf{1}.
\end{equation}
As the initial state $\rho(0)$ is known to be in the codespace, it is
clearly more beneficial to return the state $\rho(t)$ to the codespace
than to do otherwise: lacking any other information, one could at least
prepare the completely mixed state in the codespace
$\frac{1}{2}(\ketbra[\0bar,\0bar]+\ketbra[\1bar,\1bar])$, yielding an
average fidelity of $\frac{1}{2}$, rather than leaving the register
outside the codespace, yielding a fidelity of 0.  We will therefore
restrict our attention to error-correction processes $\mathcal{R}$ that
take all register states back to the codespace.  (I.e., we assume no
leakage errors during recovery.)

With the above assumption, the post-recovery state
$\mathcal{R}[\rho(t)]$ has support entirely on the codespace; thus it
can be described by its restriction to the codespace, the logical
single-qubit state $\rho_f$ such that $\Eop[\rho_f] =
\mathcal{R}[\rho(t)]$.  Call $\Dop = \Eop^\dagger \circ \mathcal{R}$ the
\textit{decoding operation} (shown to be a quantum operation in
Lemma 1 of Appendix \ref{sec:miscellaneous}):
\begin{equation}
\label{eqn:decoding_kraus}
\Dop[\rho(t)] = B^\dagger \mathcal{R}[\rho(t)] B = \sum_j B^\dagger A_j
\rho(t) A_j^\dagger B.
\end{equation}
With this definition, $\rho_f = \Dop[\rho(t)]$.  We will consider the
logical state $\rho_f$ as the outcome of the error-correction process,
and therefore may say that the code is given by its encoding and
decoding operations, i.e. $C = (\Eop,\Dop)$.  

To build intuition for the decoding operation $\Dop$, we note that for most
codes considered in the literature (and all of the specific codes
considered later in this paper) the recovery procedure is given in a
particular form.  First, a syndrome measurement is made, projecting the
register state onto one of $2^{N-1}$ orthogonal two-dimensional
subspaces; let the measurement be specified by projectors $\{P_j\}$.
After the measurement (whose outcome is given by $j$, the index of the
corresponding projector), the recovery operator $R_j$ acts on the
register, unitarily mapping the subspace projected by $P_j$ back to the
codespace.  For such codes, the recovery superoperator is given by
(\ref{eqn:recovery_kraus}) with $A_j = R_jP_j$, and
$\mathcal{R}[\rho(t)]$ is the expected state that results from averaging
over syndrome measurement outcomes.

For codes of this form, let $\{\ket[0_j], \ket[1_j]\}$ denote the
orthonormal basis for the syndrome space projected by $P_j$ such that
$R_j \ket[0_j] = \ket[\0bar]$ and $R_j \ket[1_j] = \ket[\1bar]$.  Then
$R_jP_j = \ketbra[\0bar,0_j] + \ketbra[\1bar,1_j]$, and using the
expression for $\Dop$ given in (\ref{eqn:decoding_kraus}) yields
\begin{equation}
\begin{array}{rcl}
\Dop[\rho(t)] &=& \sum_j B^\dagger R_j P_j \rho(t) P_j^\dagger
 R_j^\dagger B \\
 &=& \sum_j (\ketbra[0,0_j]+\ketbra[1,1_j]) \rho(t) 
           (\ketbra[0_j,0]+\ketbra[1_j,1])
\end{array}
\end{equation}
Thus $\rho_f = \Dop[\rho(t)]$ is the sum of the single-qubit density
matrices that result from restricting $\rho(t)$ to each of the syndrome
spaces, with basis $\{\ket[0_j], \ket[1_j]\}$ determined by the recovery
operator.

As an example, consider the bitflip code \cite{qecc_background}, a
three-qubit code that protects against single bitflip errors.  The
bitflip code's encoding transformation is given by
\begin{equation}
\label{eqn:bitflip_codewords}
\ket[0] \mapsto \ket[\0bar] = \ket[000], \;\;\;
\ket[1] \mapsto \ket[\1bar] = \ket[111].
\end{equation} 
After the action of some error dynamics, the syndrome measurement then
projects the register state into one of four subspaces: the codespace
itself, and the three subspaces that result from flipping the first,
second, or third bit of states in the codespace.  The corresponding
recovery operator simply flips the appropriate bit back, attempting to
reverse the error.  Thus the basis specifying the decoding operation is
given by
\begin{equation}
\label{eqn:bitflip_recovery_basis}
\left \{
\begin{array}{cc}
\ket[0_0] = \ket[000], & \ket[1_0] = \ket[111], \\
\ket[0_1] = \ket[100], & \ket[1_1] = \ket[011], \\
\ket[0_2] = \ket[010], & \ket[1_2] = \ket[101], \\
\ket[0_3] = \ket[001], & \ket[1_3] = \ket[110]\phantom{,} 
\end{array}
\right \}.
\end{equation}
We will use the bitflip code as an example throughout this work.

\subsection{Calculating the Effective Dynamics \label{sec:calculating_G}}

The transformation $\rho_0 \rightarrow \rho_f$ gives the effective
dynamics of the encoded information resulting from the physical dynamics
$\Nop$.  Let $\Gop$ be the map giving these effective dynamics:
$\rho_f = \Gop[\rho_0]$.  From the above discussion, the effective
dynamics are simply the result of encoding, followed by noise, followed
by decoding, i.e.
\begin{equation}
\Gop = \Dop \circ \Nop \circ \Eop.
\end{equation}
As $\Gop$ is the composition of quantum operations $\Eop$, $\Nop$ and
$\Dop$, it is itself a quantum operation.  We may therefore call $\Gop$
the \textit{effective channel} describing the code $C = (\Eop, \Dop)$
and physical noise dynamics $\Nop$.

Because the effective channel $\Gop$ is only a map on single qubit
states, it should have a compact description --- in particular, a
description much more compact than some arbitrary noise $\Nop$ acting on
$N$-qubit states.  By calculating such a compact description, we may
easily find the effective evolution of an arbitrary initial state
$\rho_0$ without explicitly considering the physical noise dynamics.  As
we now show, $\Gop$ may be written as a $4 \times 4$ matrix with a
simple interpretation.  (See \cite{Ruskai} for a full discussion of
qubit channels represented in this fashion.)

For each Pauli matrix $\sigma \in \{I,X,Y,Z\}$, let $\expval[\sigma]_0 =
\tr(\sigma \rho_0)$.  The Pauli matrices form a basis for qubit
density matrices, and so the initial logical qubit $\rho_0$ may linearly
parameterized by its expectation values $\expval[\sigma]_0$ as follows:
\begin{equation}
\label{eqn:rho_0_param}
\rho_0 = \frac{1}{2}\expval[I]_0 I + \frac{1}{2}\expval[X]_0 X +
         \frac{1}{2}\expval[Y]_0 Y + \frac{1}{2}\expval[Z]_0 Z.
\end{equation}
(As the trace of a density matrix must be 1 we will always have
$\expval[I]=1$, but it will be convenient to include this term.)
Similarly, the final logical qubit $\rho_f$ may be linearly
parameterized by its expectation values $\expval[\sigma]_f = \tr(\sigma
\rho_f)$.  Thus the effective channel $\Gop$ may be written as the
mapping from the expectation values $\expval[\sigma]_0$ of $\rho_0$ to
the expectation values $\expval[\sigma]_f$ of $\rho_f$.  Writing
$\vec{\rho}_0 = (\expval[I]_0, \expval[X]_0, \expval[Y]_0,
\expval[Z]_0)^\mathrm{T}$ and $\vec{\rho}_f = (\expval[I]_f, \expval[X]_f,
\expval[Y]_f, \expval[Z]_f)^\mathrm{T}$, the linearity of $\Gop$ allows it
to be written as the $4 \times 4$ matrix such that $\vec{\rho}_f = \Gop
\vec{\rho}_0$.  The fidelity of a pure logical qubit $\rho_0$ through
the effective channel is then given by $\tr(\rho_0 \rho_f) =
\frac{1}{2}\vec{\rho_0}^\mathrm{T}\vec{\rho_f} =
\frac{1}{2}\vec{\rho}_0^\mathrm{\ T} \Gop \vec{\rho}_0$.  Thus to fully
characterize the effective channel $\Gop$ we need only find the entries
of its $4 \times 4$ matrix representation.  (More generally, if the code
stored a $d$-dimensional state rather than the two-dimensional state of a
qubit, the logical density matrices $\rho_0$ and $\rho_f$ would be
expanded in the basis of the identity matrix and the $d^2-1$ generators
of $\mathrm{SU}(d)$, and $\Gop$ would be represented as a $d^2 \times d^2$
matrix.)

To find these matrix elements, we consider the encoding and
decoding processes in more detail.  Letting $E_\sigma$ denote
$\frac{1}{2}\Eop[\sigma]$, the encoding transformation $\Eop$ acts on
$\rho_0$ (given by (\ref{eqn:rho_0_param})) to prepare the initial
register state
\begin{equation}
\label{eqn:rho(0)_param}
\rho(0) = \expval[I]_0 E_I + \expval[X]_0 E_X +
\expval[Y]_0 E_Y + \expval[Z]_0 E_Z.
\end{equation}
Thus the encoding operation $\Eop$ is completely characterized by the
$E_\sigma$ operators, which are easily constructed from the codewords:
\begin{equation}
\label{eqn:Eops}
\begin{array}{rcl}
E_I &=& \frac{1}{2}(\ketbra[\0bar,\0bar]+\ketbra[\1bar,\1bar]) \\
E_X &=& \frac{1}{2}(\ketbra[\0bar,\1bar]+\ketbra[\1bar,\0bar]) \\
E_Y &=& \frac{1}{2}(-i\ketbra[\0bar,\1bar]+i\ketbra[\1bar,\0bar]) \\
E_Z &=& \frac{1}{2}(\ketbra[\0bar,\0bar]-\ketbra[\1bar,\1bar]).
\end{array}
\end{equation}
As expected, $\rho(0)$ is the state $\rho_0$ on the codespace, and
vanishes elsewhere.

Now consider the decoding process, which yields the logical state
$\rho_f$.  We may express the expectation values $\expval[\sigma]_f$ in
terms of $\rho(t)$, the register state prior to recovery, as follows:
\begin{equation}
\begin{array}{rcl}
\expval[\sigma]_f &=& \tr(\sigma \rho_f) = \tr(\sigma \Dop[\rho(t)]) \\
                  &=& \tr \left(\sum_j \sigma B^\dagger A_j \rho(t)
                                       A_j^\dagger B \right ).
\end{array}
\end{equation}
Exploiting the cyclic property of the trace and noting that $B \sigma
B^\dagger = \Eop[\sigma] = 2E_\sigma$, we have
\begin{equation}
\label{eqn:sigma_f}
\expval[\sigma]_f = \tr(D_\sigma \rho(t))
\;\;\;\mathrm{where}\;\;\;
D_\sigma = 2\sum_j A_j^\dagger E_\sigma A_j.
\end{equation}
Thus the decoding operation $\Dop$ is completely characterized by the
$D_\sigma$ operators.  

Substituting $\rho(t) = \Nop[\rho(0)]$ into (\ref{eqn:sigma_f}), we have
$\expval[\sigma]_f = \tr(D_\sigma \Nop[\rho(0)])$.  Substituting in the
expression for $\rho(0)$ given by (\ref{eqn:rho(0)_param}) then yields
\begin{equation}
\expval[\sigma]_f = \tr \left ( D_\sigma\Nop \left [ \sum_{\sigma'}
                        \expval[\sigma']_0 E_{\sigma'} \right ] \right
                        ).
\end{equation}
Letting the matrix elements of $\Gop$ be given by
\begin{equation}
\label{eqn:G_formula}
\Gop_{\sigma \sigma'} = \tr ( D_\sigma \Nop 
                           \left [ E_{\sigma'} \right ])
\end{equation}
for $\sigma, \sigma' \in \{I,X,Y,Z\}$, we have $\expval[\sigma]_f =
\sum_{\sigma'}G_{\sigma\sigma'}\expval[\sigma']_0$, i.e.  $\vec{\rho}_f
= \Gop \vec{\rho}_0$. 

To completely characterize the effective channel $\Gop$, then, we need
only compute these matrix elements.  In fact, trace-preservation
(i.e. $\expval[I]_f = \expval[I]_0)$ requires $\Gop_{II}=1$ and
$\Gop_{IX} = \Gop_{IY} = \Gop_{IZ} = 0$.  Thus the effect on the logical
information of the potentially complex dynamics of the $N$-qubit
register space are characterized by the remaining twelve matrix elements
of $\Gop$.  If $\Nop$ is time-dependent, then the only observable
effects of this time-dependence will appear in the time dependence of
the $\Gop_{\sigma \sigma'}$, and $\Gop_t$ gives the effective channel
for correction performed at time $t$.  Note that the dynamics $\Nop$
need not be related to those against which the code was designed to
protect.

We have thus shown that the effective dynamics may be calculated by
evaluating (\ref{eqn:G_formula}), which requires constructing the
$E_\sigma$ and $D_\sigma$ operators.  The $E_\sigma$ operators are
easily understood to be the operators which act as $\frac{1}{2}\sigma$
on the codespace and vanish elsewhere; to build intuition for the
$D_\sigma$ operators, consider codes whose recovery is specified by 
syndrome measurement projectors $\{P_j\}$ and recovery operators
$\{R_j\}$ as discussed in section \ref{sec:qec_basics}.  For these codes, we
have $A_j = R_jP_j$, and so $D_\sigma = 2\sum_j P_j^\dagger R_j^\dagger
E_\sigma R_j P_j$.  This expression may be simplified by noting that
$E_\sigma$ maps the codespace to itself and vanishes elsewhere, and
$R_j$ unitarily maps the space projected by $P_j$ to the codespace.
Thus $R_j^\dagger E_\sigma R_j$ unitarily maps the space projected by
$P_j$ to itself and vanishes elsewhere, i.e. $P_j R_j^\dagger E_\sigma
R_j P_j = R_j^\dagger E_\sigma R_j$.  We therefore have
\begin{equation}
\label{eqn:Dops_nobasis}
D_\sigma = 2\sum_j R_j^\dagger E_\sigma R_j.
\end{equation}
Using the expressions for $E_\sigma$ given in (\ref{eqn:Eops}) and $R_j =
\ketbra[\0bar,0_j]+\ketbra[\1bar,1_j]$, we have
\begin{equation}
\label{eqn:Dops}
\begin{array}{rcl}
D_I &=& \sum_j (\ketbra[0_j,0_j]+\ketbra[1_j,1_j]) \\
D_X &=& \sum_j (\ketbra[0_j,1_j]+\ketbra[1_j,0_j]) \\
D_Y &=& \sum_j (-i\ketbra[0_j,1_j]+i\ketbra[1_j,0_j]) \\
D_Z &=& \sum_j (\ketbra[0_j,0_j]-\ketbra[1_j,1_j]). 
\end{array}
\end{equation}
Thus we see that in this case $D_\sigma$ is simply the sum of the
operators $\sigma$ acting on each of the syndrome spaces, with
$Z$-eigenstates $\ket[0_j]$ and $\ket[1_j]$ determined by the recovery
procedure.  Note that $D_I$ is the identity operator on
the entire register space.

\section{Coding as a Map on Channels\label{sec:coding_map}}

One often considers noise models $\Nop$ consisting of uncorrelated noise
on each of the $N$ physical qubits.  This type of model arises naturally
in a communication setting, where the register qubits are sent over a
noisy transmission line one at a time, and is also appropriate for various
physical implementations of a quantum computer.  (By contrast, one can
also consider error models in which correlated noise dominates
\cite{collective}.)  For such models, we may write
\begin{equation}
\label{eqn:uncorrelated_noise}
\Nop = \Nop^{(1)}\otimes\Nop^{(1)}\otimes
\ldots\otimes\Nop^{(1)} = \Nop^{(1)\otimes N}
\end{equation}
where $\Nop^{(1)}$ is a quantum operation on a single qubit.

The goal of encoding a qubit is to suppress decoherence: multiple qubits
are employed to yield an effective channel $\Gop$, which should be less
noisy than the channel resulting from storing information in a single
physical qubit, namely $\Nop^{(1)}$.  A code can thus be seen as a map
on channels, taking $\Nop^{(1)}$ to $\Gop$.  More precisely, for an
$N$-qubit code $C = (\Eop, \Dop)$, define the corresponding \textit{coding map}
$\Omega^{C}$ by
\begin{equation}
\label{eqn:omega_definition}
\Omega^{C}:\Nop^{(1)} \rightarrow \Gop = \Dop \circ \Nop^{(1)\otimes N}
\circ \Eop.
\end{equation}

We now derive an expression for the coding map $\Omega^{C}$ of an
arbitrary code $C = (\Eop,\Dop)$.  In section
\ref{sec:calculating_G} we described how $\Gop$ may be specified by its
matrix elements $\Gop_{\sigma \sigma'}$, given by (\ref{eqn:G_formula}).
Since $\Nop^{(1)}$ is a single-qubit quantum operation, it may also be
written as a $4 \times 4$ matrix such that if $\Nop^{(1)}$ takes $\rho$
to $\rho'$, then $\vec{\rho}' = \Nop^{(1)}\vec{\rho}$ .  We seek an
expression for the matrix elements of the effective channel $\Gop$ in
terms of the matrix elements of the physical channel $\Nop^{(1)}$.

Operators on $N$ qubits may be written as sums of tensor products of $N$
Pauli matrices; we may therefore write the $E_{\sigma}$ and $D_\sigma$
operators describing $C = (\Eop, \Dop)$ as
\begin{eqnarray}
\label{eqn:E_pauli}
E_{\sigma'} &=& 
\sum_{\genfrac{}{}{0pt}{}{\mu_i \in}{\{I,X,Y,Z\}}}
\alpha^{\sigma'}_{\{\mu_i\}}
\left (\tfrac{1}{2} \mu_1 \right) 
\otimes \ldots \otimes 
\left (\tfrac{1}{2} \mu_N \right) 
\\
\label{eqn:D_pauli}
D_\sigma &=& \sum_{ 
\genfrac{}{}{0pt}{}{\nu_i \in}{\{I,X,Y,Z\}}}
\beta^\sigma_{\{\nu_i\}}
\nu_1 \otimes \ldots \otimes \nu_N.
\end{eqnarray}
E.g., for the bitflip code described in section \ref{sec:qec_basics} by 
(\ref{eqn:bitflip_codewords}) and (\ref{eqn:bitflip_recovery_basis}), we
may calculate the $E_{\sigma'}$ and $D_\sigma$ operators using
(\ref{eqn:Eops}) and (\ref{eqn:Dops}); expanding the results in the basis of
Pauli operators yields
\begin{equation}
\label{eqn:bitflip_Eops}
{\setlength{\arraycolsep}{2pt}
\begin{array}{rclcccccccl}
E_I &=&\frac{1}{8}( & III  & + & IZZ & + & ZIZ & + & ZZI & ) \\ 
E_X &=&\frac{1}{8}( & XXX  & - & XYY & - & YXY & - & YYX & ) \\
E_Y &=&\frac{1}{8}( & -YYY & + & YXX & + & XYX & + & XXY & ) \\
E_Z &=&\frac{1}{8}( & ZZZ  & + & ZII & + & IZI & + & IIZ & )
\end{array}}
\end{equation}
and
\begin{equation}
\label{eqn:bitflip_Dops}
{\setlength{\arraycolsep}{2pt}
\begin{array}{ccrcccccccl}
D_I &=& & III &   \\
D_X &=& & XXX \\
D_Y &=& \frac{1}{2}( & YYY & + & YXX & + & XYX & + & XXY &) \\
D_Z &=& \frac{1}{2}( & -ZZZ & + & ZII & + & IZI & + & IIZ &).
\end{array}}
\end{equation}

To find the matrix elements of the effective channel, we substitute
(\ref{eqn:uncorrelated_noise}), (\ref{eqn:E_pauli}), and
(\ref{eqn:D_pauli}) into the expression for these matrix elements given
by (\ref{eqn:G_formula}).  Noting that $\Nop[\frac{1}{2}\mu_1 \otimes
\ldots \otimes \frac{1}{2}\mu_N] = \Nop^{(1)}[\frac{1}{2}\mu_1] \otimes
\ldots \otimes \Nop^{(1)}[\frac{1}{2}\mu_N]$ and $\tr((A \otimes B)(C
\otimes D)) = \tr(AC)\tr(BD)$ yields
\begin{equation}
\Gop_{\sigma \sigma'} = 
\sum_{\{\mu_i\},\{\nu_i\}} 
\left (
\beta_{\{\nu_i\}}^{\sigma}
\alpha_{\{\mu_i\}}^{\sigma'} \prod_{i = 1}^N 
\tr(\nu_i \Nop^{(1)}[\tfrac{1}{2} \mu_i])
\right ).
\end{equation}
From the orthogonality of Pauli matrices, the matrix $\frac{1}{2}\mu_i$,
when written as a vector of expectation values, has a 1 in the $\mu_i$
component and zeros elsewhere.  Further, $\tr(\nu_i \rho)$ is simply the
$\nu_i$ component of $\vec{\rho}$.  Thus $\tr(\nu_i
\Nop^{(1)}[\tfrac{1}{2} \mu_i]) = \Nop^{(1)}_{\nu_i \mu_i}$, and we have
\begin{equation}
\label{eqn:G_uncorrelated}
\Gop_{\sigma \sigma'} =
\sum_{\{\mu_i\},\{\nu_i\}}
\left (
\beta_{\{\nu_i\}}^{\sigma}
\alpha_{\{\mu_i\}}^{\sigma'}
\prod_{i = 1}^N \Nop^{(1)}_{{\nu_i}{\mu_i}} \right ).
\end{equation}

Thus the matrix elements of $\Gop$ can be expressed as polynomials of
the matrix elements of $\Nop^{(1)}$, with the polynomial coefficients
depending only on the $E_{\sigma'}$ and $D_\sigma$ of the code.  These
polynomials specify $\Omega^C$.  By computing these polynomials for a
code $C$, one can easily calculate the effective channel for the code
$C$ due to any error model with identical, uncorrelated noise acting on
each physical qubit.  (If a different noise model acts on each physical
qubit, i.e. $\Nop = \Nop^{(1)} \otimes \ldots \otimes \Nop^{(N)}$,
simply replace $\Nop^{(1)}_{{\nu_i}{\mu_i}}$ with
$\Nop^{(i)}_{{\nu_i}{\mu_i}}$ in (\ref{eqn:G_uncorrelated}).)

\section{Concatenated Codes\label{sec:concatenated}}

We now consider concatenated codes \cite{qecc_background,
concatenation_original}. We first describe the procedure for constructing
such codes, and then show how the coding maps $\Omega^C$ make the
calculation of the effective channels for such codes straightforward.

\subsection{Constructing Concatenated Codes}

\begin{figure*}
\includegraphics{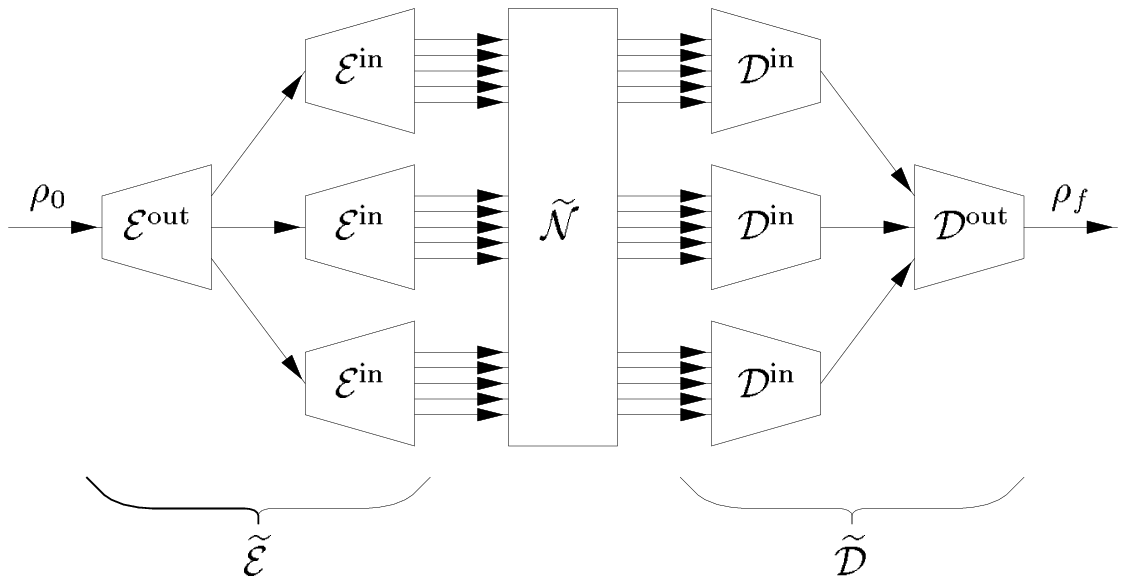}
\caption{The error-correction process for the concatenated code
$C^\out(C^\inn) = (\wt{\Eop},\wt{\Dop})$; here $C^\out =
(\Eop^\out,\Dop^\out)$ is a three-qubit code and $C^\inn =
(\Eop^\inn,\Dop^\inn)$ is a five-qubit code.  The noise process
$\widetilde{\Nop}$ acts on the entire 15-qubit register.
\label{fig:concatenation}}
\end{figure*}

We now describe how two codes may be concatenated to form a larger code;
the procedure is depicted in Figure \ref{fig:concatenation}.  Let the
two codes be an $M$-qubit code $C^\out = (\Eop^\out, \Dop^\out)$, called
the outer code, and an $N$-qubit code $C^\inn = (\Eop^\inn, \Dop^\inn)$,
called the inner code.  A logical qubit $\rho_0$ is encoded first using
the outer code $C^\out$, yielding the $M$-qubit state
$\Eop^\out[\rho_0]$.  Each of these qubits is then encoded by the inner
code; i.e., the map $\Eop^\inn\otimes\ldots\otimes\Eop^\inn =
(\Eop^\inn)^{\otimes M}$ acts on $\Eop^\out[\rho_0]$.  The composition
of these encodings forms the encoding map for the concatenated code:
\begin{equation}
\label{eqn:Etilde}
\wt{\Eop} = 
(\Eop^\inn)^{\otimes M} \circ \Eop^\out.
\end{equation}
The $M$ sections of the register encoding each of the $M$ qubits in
$\Eop^\out[\rho_0]$ are called blocks; each block contains $N$ qubits.
After the encoding, a noise process $\widetilde{\Nop}$ acts on the
entire $MN$-qubit register.

A simple error-correction scheme (and one that seems
reasonable for use in a scalable architecture) coherently corrects each
of the code blocks based on the inner code, and then corrects the entire
register based on the outer code.  I.e., the decoding map for the
concatenated code is given by
\begin{equation}
\label{eqn:Dtilde}
\wt{\Dop} = 
\Dop^\out \circ \Dop^{\inn\otimes M}.
\end{equation}
We denote the concatenated code (with this correction scheme) by
$C^\out(C^\inn) = (\wt{\Eop},\wt{\Dop})$; note that $C^\out(C^\inn)$ is
an $MN$-qubit code.  

\subsection{Effective Channels for Concatenated Codes}

Suppose that we have computed the effective channel $\Gop$ due to a code
$C^\inn = (\Eop^\inn, \Dop^\inn)$ with some noise dynamics $\Nop$, and wish
to consider the effective channel $\wt{\Gop}$ resulting from the
concatenated code $C^\out(C^\inn)$.  We assume that each $N$-bit block in
the register evolves according to the noise dynamics $\Nop$ and no
cross-block correlations are introduced, i.e. that the evolution
operator on the $MN$-bit register is
\begin{equation}
\label{eqn:Ntilde}
\wt{\Nop} = \Nop \otimes \Nop \otimes
\ldots \otimes \Nop = \Nop^{\otimes M}.
\end{equation}
By definition, we have $\wt{\Gop}$ =
$\wt{\Dop} \circ \wt{\Nop} \circ \wt{\Eop}$.  Substituting
(\ref{eqn:Etilde}), (\ref{eqn:Dtilde}) and (\ref{eqn:Ntilde}) into this
expression yields
\begin{equation}
\begin{array}{rcl}
\wt{\Gop} & = &\Dop^\out \circ \Dop^{\inn \otimes M} \circ
                  \Nop^{\otimes M} \circ \Eop^{\inn \otimes M} \circ
                  \Eop^\out \\
             & = &\Dop^\out \circ (\Dop^\inn \circ \Nop \circ
                  \Eop^\inn)^{\otimes M} \circ \Eop^\out \\
             & = &\Dop^\out \circ \Gop^{\otimes M} \circ \Eop^\out
\end{array}
\end{equation}
where we have used $\Gop = \Dop^\inn \circ \Nop \circ \Eop^\inn$.  This
result makes sense: each of the $M$ blocks of $N$ bits represents a
single logical qubit encoded in $C^\inn$, and as the block has dynamics
$\Nop$, this logical qubit's evolution will be described by $\Gop$.
Comparing with the definition of the coding map
(\ref{eqn:omega_definition}), we then have
\begin{equation}
\wt{\Gop} = \Omega^{C^\out}(\Gop).
\end{equation}
Thus given the effective channel for a code $C^\inn$ and an error model, 
the coding map $\Omega^{C^\out}$ makes it straightforward to
compute the effective channel due to the concatenated code
$C^\out(C^\inn)$.

Further, suppose that the original noise model $\Nop$ had the form of
uncorrelated noise on single physical qubits, as given by
(\ref{eqn:uncorrelated_noise}).  Then $\Gop =
\Omega^{C^\inn}(\Nop^{(1)})$, and so $\wt{\Gop} = \Omega^{C^\out}(
\Omega^{C^\inn}(\Nop^{(1)}))$.  We may therefore conclude that composing
coding maps gives the coding map for the concatenated code, i.e.
\begin{equation}
\Omega^{C^\out(C^\inn)} = \Omega^{C^\out} \circ \Omega^{C^\inn}.
\end{equation}

More generally, we may characterize both the finite and asymptotic
behavior of any concatenation scheme involving the codes $\{C_k\}$ by
computing the maps $\Omega^{C_k}$.  Then the finite concatenation scheme
$C_1(C_2(\ldots C_n\ldots))$ is characterized by $\Omega^{C_1(C_2(\ldots
C_n\ldots))} = \Omega^{C_1}\circ\Omega^{C_2}\circ \ldots \circ \Omega^{C_n}$.
We expect the typical $\Omega^C$ to be sufficiently well-behaved that
standard dynamical systems methods \cite{dynamical_standard} will yield
the $\ell \rightarrow \infty$ limit of $(\Omega^C)^\ell$; one need not
compose the $(\Omega^C)^\ell$ explicitly.  In section \ref{sec:thresholds},
we will consider such asymptotic limits in more detail.

\section{Diagonal Channels\label{sec:diagonal}}
As an application of the methods presented above, we will consider the
commonly-considered error model in which each physical register qubit is
subjected to the symmetric depolarizing channel \cite{qecc_background}.
These single-qubit noise dynamics are given by the master equation
\begin{equation}
\frac{d\rho}{dt} = 
\frac{\gamma}{4}\mathcal{L}_X[\rho] +
\frac{\gamma}{4}\mathcal{L}_Y[\rho] +
\frac{\gamma}{4}\mathcal{L}_Z[\rho]
\end{equation}
where for any linear qubit operator $c$
the Lindblad decoherence operator $\mathcal{L}_c$ is given by
\begin{equation}
\mathcal{L}_c[\rho] = c \rho c^\dagger - \frac{1}{2}c^\dagger c \rho
                - \frac{1}{2}\rho c^\dagger c
\end{equation}
and $\gamma$ is a measure of the noise strength.  This master equation
is easily solved, yielding a qubit channel with matrix representation
\begin{equation}
\label{eqn:depolarizing_matrix}
\Ndep_t=
\left (
\begin{array}{cccc}
1 & 0 & 0 & 0 \\
0 & e^{-\gamma t} & 0 & 0 \\
0 & 0 & e^{-\gamma t} & 0 \\
0 & 0 & 0 & e^{-\gamma t}
\end{array}
\right ).
\end{equation}
Before calculating effective channels due to this error model, it will
be useful to discuss the more general set of channels whose matrix
representation is diagonal.  As we will see, these channels correspond
to single-bit Pauli channels, and will allow us to demonstrate the power
of the techniques developed above.

Consider a qubit channel given by a diagonal matrix $\Nop^{(1)}$.  From
trace preservation $\Nop^{(1)}_{II}=1$, so let the channel with
$\Nop^{(1)}_{XX}=x$, $\Nop^{(1)}_{YY}=y$ and $\Nop^{(1)}_{ZZ}=z$ be
denoted $[x,y,z]$ for compactness.  (Thus the depolarizing channel
(\ref{eqn:depolarizing_matrix}) is given by $[e^{-\gamma t}, e^{-\gamma
t},e^{-\gamma t}]$.)  In \cite{King} it is shown that complete
positivity of such a channel requires
\begin{equation}
\label{eqn:CP_constraints}
\begin{array}{rl}
-x+y+z & \leq 1 \\
 x-y+z & \leq 1 \\
 x+y-z & \leq 1 \\
-x-y-z & \leq 1.
\end{array}
\end{equation}

Now consider the single-bit Pauli channel in which the transmitted state
is subjected to the Pauli operators $X$, $Y$, and $Z$ with
exclusive probabilities $p_X$, $p_Y$, and $p_Z$, i.e.
\begin{eqnarray}
\label{eqn:pauli_channel}
\rho &\rightarrow& (1-p_X-p_Y-p_Z)\rho \nonumber \\
  &&       + p_X X\rho X + p_Y Y\rho Y + p_Z Z \rho Z.
\end{eqnarray}
It is easy to show that this channel has the diagonal matrix
representation 
\begin{equation}
[1-2(p_Y+p_Z),\: 1-2(p_X+p_Z),\: 1-2(p_X+p_Y)],
\end{equation}
and so any Pauli channel is a diagonal channel.  The converse is also
true: choosing $p_x = \frac{1+x-y-z}{4}$, $p_y = \frac{1-x+y-z}{4}$, and
$p_z = \frac{1-x-y+z}{4}$ yields the channel $[x,y,z]$, and the complete
positivity constraints (\ref{eqn:CP_constraints}) yield the standard
probability rules $p_X,p_Y,p_Z \geq 0$ and $p_X+p_Y+p_Z \leq 1$.  Thus
any diagonal channel may be realized as a Pauli channel.  Pauli channels
are among the most commonly considered error models in the literature,
and we will restrict our attention to diagonal channels for the
remainder of this work.

The effect of a diagonal channel on a qubit is simple to interpret: we
have $\expval[X]_f = x\expval[X]_0$, $\expval[Y]_f = y\expval[Y]_0$, and
$\expval[Z]_f = z\expval[Z]_0$.  Thus the $X$, $Y$,
and $Z$ components of $\vec{\rho}_0$ decay independently, and
we may therefore speak of the decoherence of $\expval[X]$, $\expval[Y]$,
and $\expval[Z]$.  Recalling from section \ref{sec:calculating_G} that
the fidelity of a pure state $\rho$ through a qubit channel $\Gop$ is
given by $\frac{1}{2}\vec{\rho}^\mathrm{T}\Gop\vec{\rho}$, the
respective fidelities of $X$-, $Y$-, and $Z$-eigenstates through the
channel are $\frac{1}{2}(1+x)$, $\frac{1}{2}(1+y)$, and
$\frac{1}{2}(1+z)$.  More generally, the fidelity of a pure state
(requiring $\expval[X]^2+\expval[Y]^2+\expval[Z]^2=1$) is given by
$\frac{1}{2}(1+x\expval[X]^2+y\expval[Y]^2+z\expval[Z]^2)$.  A common
figure of merit for a channel is the worst-case fidelity of a pure
state, which for a diagonal channel is $\frac{1}{2}(1+\min(x,y,z))$.
Thus if for a given error model a code $C$ yields an effective channel
$[x,y,z]$ and a code $C'$ yields an effective channel $[x',y',z']$, we
say that $C$ outperforms $C'$ if $\min(x,y,z) > \min(x',y',z')$.

Many commonly considered codes are \textit{stabilizer codes}
\cite{qecc_background,Gottesman}, which are designed to detect and
correct Pauli errors; it would therefore not be so surprising if the
coding maps for such codes were particularly well-behaved when acting on
a Pauli channel.  In fact, as proved in Appendix \ref{sec:stabilizer},
if $C$ is a stabilizer code and $\Nop^{(1)}$ is diagonal, then
$\Omega^C(\Nop^{(1)})$ is also diagonal.  Thus just as arbitrary codes
act as maps on the space of qubit channels, stabilizer codes act as
maps on the space of diagonal qubit channels.

\section{Exact Performance for Several Codes of 
Interest\label{sec:performance}}

We will now present the effective channels for several codes of
interest under diagonal error models.  The codes considered here may
all be formulated as stabilizer codes; thus, as described in the previous
section, the effective channels will also be diagonal.  The diagonal
elements of the effective channel $\Gop = \Omega^C([x,y,z])$ may be
calculated either using the coding map methods presented in section
\ref{sec:coding_map}, or using the stabilizer formalism as shown in
Appendix \ref{sec:stabilizer}, which may be computationally
advantageous.  For each code, we will compute the effective channel for
a general diagonal error model $[x,y,z]$, and then interpret the results
for the symmetric depolarizing channel $\Ndep_t = [e^{-\gamma
t},e^{-\gamma t},e^{-\gamma t}]$.

The bitflip code first mentioned in section \ref{sec:qec_basics} is a
stabilizer code; letting $\Omega^\mathrm{bf}$ denote the corresponding
coding map, we find
\begin{equation}
\label{eqn:poly_bit}
\Omega^\btf([x,y,z])=
\left [x^3,\: \tfrac{3}{2}x^2 y-\tfrac{1}{2}y^3, 
            \:  \tfrac{3}{2}z-\tfrac{1}{2}z^3 \right ].
\end{equation}
As the bitflip code is only a three-qubit code, it is not unreasonable
to check this result with more conventional methods, e.g. by counting
bitflip and phaseflip errors, or by working in the Heisenberg picture to
compute the evolution of the relevant expectation values.  However, for
larger codes such computations will rapidly become unmanageable.

\begin{figure}
\includegraphics[scale=0.47]{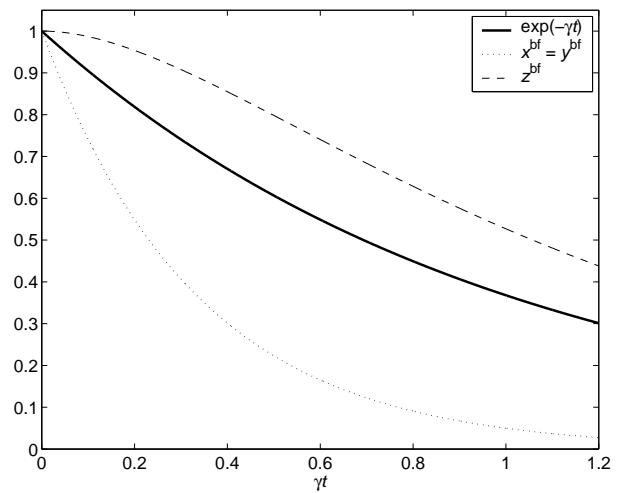}
\caption{The effective channel $[x^\btf(t),y^\btf(t),z^\btf(t)]$ due to
the bitflip code under the symmetric depolarizing channel.  The
respective fidelities of $X$, $Y$ and $Z$ eigenstates for correction
performed at time $t$ are given by $\frac{1}{2}(1+x^\btf(t))$,
$\frac{1}{2}(1+y^\btf(t))$, and $\frac{1}{2}(1+z^\btf(t))$.
\label{fig:bitflip_performance}}
\end{figure}

To examine the bitflip code acting under the symmetric depolarizing
channel, define $[x^\btf(t),y^\btf(t),z^\btf(t)] =
\Omega^\btf(\Ndep_t)$; the functions $x^\btf$, $y^\btf$, and $z^\btf$
are plotted in Figure \ref{fig:bitflip_performance} along with $e^{-\gamma
t}$ (describing the decoherence of the physical qubits) for comparison.
We see that $z^\btf(t) > e^{-\gamma t}$, and thus the decoherence of
$\expval[Z]$ is suppressed by the bitflip code.  However, $x^\btf(t) =
y^\btf(t) < e^{-\gamma t}$, and thus the decoherences of $\expval[X]$
and $\expval[Y]$ are increased by the bitflip code.

More generally, for any $0 < x,z < 1$ we have
$\frac{3}{2}z-\frac{1}{2}z^3 > z$ and $x^3 < x$, so for any physical
channel in this regime the bitflip code always suppresses decoherence of
$\expval[Z]$ and increases decoherence of $\expval[X]$.  Decoherence of
$\expval[Y]$ is suppressed when $x > \sqrt{\frac{2}{3}}$ and $0 < y <
\sqrt{3x^2 - 2}$, and increased for all other positive values of $x$ and
$y$.  We may therefore conclude that under a general Pauli channel the
bitflip code increases the fidelity of some transmitted states at the
expense of others, and thus the bitflip code is outperformed by storing
the logical qubit in a single physical bit.

However, as the bitflip code is designed to only protect against
physical bitflip ($X$) errors, it should not be expected to perform well
in the presence of arbitrary Pauli errors.  If we consider physical
channels with only $X$ errors, we find that the bitflip code suppresses
decoherence of all encoded states.  More precisely, suppose that the
physical qubits are evolving via a Pauli channel
(\ref{eqn:pauli_channel}) with only $X$ errors, i.e. $p_Y = p_Z = 0$.
Then $[x,y,z] = [1,1-2p_X,1-2p_X]$, and $\Omega^\btf([x,y,z]) =
[1,1-\frac{3}{2}p_X^2 +\frac{1}{2}p_X^3, 1-\frac{3}{2}p_X^2
+\frac{1}{2}p_X^3]$.  Thus we have reproduced the usual result of a
leading-order analysis: the bitflip code suppresses decoherence due to
$X$ errors to order $p_X^2$.

Now consider the three-qubit phaseflip code \cite{qecc_background}, with
encoding $\ket[\pm] \mapsto \ket[\pm\pm\pm]$ for $\ket[\pm] =
\frac{1}{\sqrt{2}}(\ket[0]+\ket[1])$.  This code is completely analogous
to the bitflip code, detecting and correcting single phaseflip ($Z$)
errors instead of single bitflip ($X$) errors.  The phaseflip code's
coding map $\Omega^\phf$ is exactly the same as that of the bitflip
code, with the role of $X$ and $Z$ interchanged:
\begin{equation}
\label{eqn:poly_phase}
\Omega^\phf([x,y,z])= \left [\tfrac{3}{2}x-\tfrac{1}{2}x^3,\:
\tfrac{3}{2}z^2 y-\tfrac{1}{2}y^3, \: z^3 \right ].
\end{equation}
The concatenation phaseflip(bitflip) yields the Shor nine-bit code
\cite{qecc_original, qecc_background} with encoding $\ket[\pm] \mapsto
\frac{1}{\sqrt{8}}(\ket[000] \pm \ket[111])^{\otimes 3}$.  Thus
$\Omega^\Shor=\Omega^\phf \circ \Omega^\btf$.  Evaluating this
composition using the coding maps (\ref{eqn:poly_bit}) and
(\ref{eqn:poly_phase}),
\begin{eqnarray}
\label{eqn:form_poly_shor}
\Omega^\Shor([x,y,z]) &=& \Omega^\phf(\Omega^\btf([x,y,z])) \nonumber \\
&=&[P(x),Q(x,y,z),R(z)]
\end{eqnarray}
where
\begin{equation}
\label{eqn:poly_shor}
\begin{array}{rcl}
P(x)&=& \tfrac{3}{2}x^3 - \tfrac{1}{2}x^9  \\
Q(x,y,z) &=&  \tfrac{3}{2}\left ( \tfrac{3}{2}z - \tfrac{1}{2}z^3 \right )^2
  \left (\tfrac{3}{2}x^2 y - \tfrac{1}{2}y^3 \right ) \\
& & -\tfrac{1}{2}\left (\tfrac{3}{2}x^2 y - \tfrac{1}{2}y^3 \right
   )^3 \\
R(z) &=& \left (\tfrac{3}{2}z - \tfrac{1}{2}z^3 \right )^3. 
\end{array}
\end{equation}
(The combinatoric analysis required to reproduce this result by counting
bitflip and phaseflip errors would be quite tedious!)  

\begin{figure}
\includegraphics[scale=0.47]{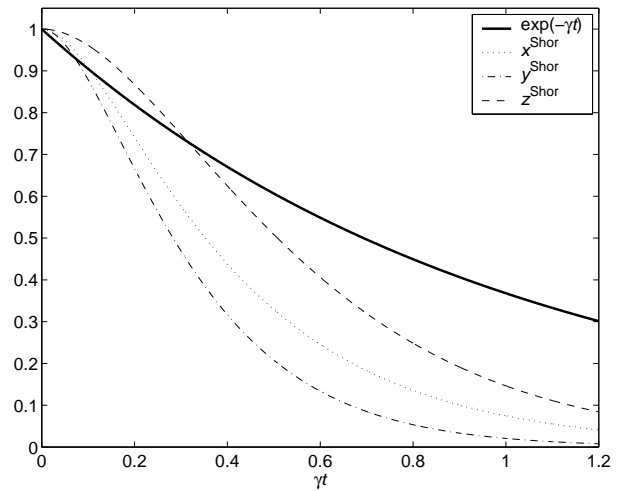}
\caption{The effective channel $[x^\Shor(t),y^\Shor(t),z^\Shor(t)]$ due to
the Shor code under the symmetric depolarizing channel.  
\label{fig:shor_performance}}
\end{figure}

To examine the Shor code acting on the symmetric depolarizing channel,
let $[x^\Shor(t),y^\Shor(t),z^\Shor(t)] = \Omega^\Shor(\Ndep_t)$; the
functions $x^\Shor$, $y^\Shor$ and $z^\Shor$ are plotted in
Figure \ref{fig:shor_performance}.  We see that for short times (or
equivalently, weak noise-strength $\gamma$), the Shor code suppresses
decoherence of $\expval[X]$, $\expval[Y]$, and $\expval[Z]$.  For long
times, however, the code increases the decoherence of all three
expectation values, and as $z^\Shor(t) > x^\Shor(t) > y^\Shor(t)$, in 
an intermediate regime the code suppresses the decoherence of some of
the expectation values while increasing that of others.  Thus to
suppress the decoherence of an arbitrary logical state, correction needs
to be performed at a time $t$ when $y^\Shor(t) >
e^{-\gamma t}$.

Above we defined the phaseflip code by the encoding $\ket[\pm] \mapsto
\ket[\pm\pm\pm]$; we could have also used the encoding given by $\ket[0]
\mapsto \ket[+++]$, $\ket[1]\mapsto \ket[---]$.  Call the code with this
encoding phaseflip$'$, with coding map $\Omega^{\phf'}$.  As this
modification of the phaseflip code simply interchanges the encoded $X$-
and $Z$-eigenstates, the new effective channel is simply that of the
original phaseflip code with the effects of the channel on the $X$ and
$Z$ components of $\vec{\rho}_0$ interchanged: $\Omega^{\phf '}([x,y,z])
= [z^3,\frac{3}{2}z^2 y-\frac{1}{2}y^3, \frac{3}{2}x-\frac{1}{2}x^3]$
(compare to (\ref{eqn:poly_phase})).  We could then use this version of
the phaseflip code to define an alternative version of the Shor code
with the encoding $\ket[0] \mapsto \frac{1}{\sqrt{8}}(\ket[000] +
\ket[111])^{\otimes 3}$, $\ket[1] \mapsto \frac{1}{\sqrt{8}}(\ket[000] -
\ket[111])^{\otimes 3}$.  Call this code Shor$'$, with corresponding
coding map $\Omega^{\Shor'} = \Omega^{\phf'}\circ \Omega^\btf$.  We find
\begin{equation}
\label{eqn:form_poly_shor_prime}
\Omega^{\Shor'}([x,y,z]) = [R(z),Q(x,y,z),P(x)].
\end{equation}
with the polynomials $P$, $Q$ and $R$ defined by (\ref{eqn:poly_shor}).
Comparing to (\ref{eqn:form_poly_shor}), we see that again this
modification of the Shor code simply interchanges the effect of the
channel on $X$ and $Z$ components of $\vec{\rho}_0$.  Assuming that the
encoded logical states are randomly distributed (as opposed to always
sending $Z$-eigenstates, for example), the choice of using the Shor code
or the Shor$'$ code is simply one of aesthetics: the effective channels
are identical up to interchange of the decoherence of $\expval[X]$ and
$\expval[Z]$.  However, as we will see in the next section, this choice
does have an impact when these codes are concatenated.

For comparison, we consider two other stabilizer codes of interest.  The
Steane code \cite{qecc_background, steane} is a seven-bit code designed
to correct errors consisting either of a Pauli error ($X$, $Y$, or $Z$)
on a single qubit of the codeword, or of an $X$ and a $Z$ error on
separate qubits.  We find
\begin{equation}
\label{eqn:poly_steane}
\Omega^\mathrm{Steane}([x,y,z]) = [S(x),T(x,y,z),S(z)]
\end{equation}
with
\begin{equation}
\begin{array}{rcl}
S(x) &=& \tfrac{7}{4}x^3 - \tfrac{3}{4}x^7 \\
T(x,y,z)&=& \tfrac{7}{16}y^3 + \tfrac{9}{16}y^7
-\tfrac{21}{16}(x^4+z^4)y^3  + \tfrac{21}{8}x^2yz^2.
\end{array}
\end{equation}
Let $[x^\Steane(t),y^\Steane(t),z^\Steane(t)] =
\Omega^\Steane(\Ndep_t)$; we find that the functions $x^\Steane$,
$y^\Steane$ and $z^\Steane$ are qualitatively similar to the
analogous functions of the Shor code.  If they were plotted in
Figure \ref{fig:shor_performance}, these functions would be interspersed
between the plotted curves: for all values of $t > 0$, we have $z^\Shor
> z^\Steane = x^\Steane > x^\Shor > y^\Steane > y^\Shor$.  Though the
Shor code more effectively suppresses decoherence for logical
$Z$-eigenstates, the Steane code performs better in the worst case
($Y$-eigenstates), and thus outperforms the Shor code.

The Five-Bit code \cite{qecc_background, laflamme_miqel,
bennett_divincenzo} corrects Pauli errors on a single qubit of the
codeword.  We find
\begin{equation}
\label{eqn:poly_five}
\Omega^\mathrm{Five}([x,y,z]) = 
[U(x,y,z),U(y,z,x),U(z,x,y)] 
\end{equation}
with
\begin{equation}
U(x,y,z) = 
\tfrac{5}{4}x(y^2+z^2)-\tfrac{5}{4}xy^2z^2 -\tfrac{1}{4}x^5.
\end{equation}
Letting $[x^\Five(t),y^\Five(t),z^\Five(t)] = \Omega^\Five(\Ndep_t)$
yields $x^\Five = y^\Five = z^\Five$, as expected from the symmetries of
the code and of the map $\Omega^\Five$.  Thus the fidelity of a state
through this channel is independent of the state.  These functions also
have qualitatively similar behavior to those plotted in
Figure \ref{fig:shor_performance}, and for $t > 0$ we have $z^\Shor >
z^\Five > z^\Steane > x^\Shor$.  Thus the Five-Bit code
outperforms both the Shor and Steane codes.

\section{Exact Performance and Thresholds for Certain Concatenation Schemes\label{sec:thresholds}}

We now consider the effective channel due to families of concatenated
codes under the symmetric depolarizing channel.  First, consider the
Shor code concatenated with itself $\ell$ times, denoted by Shor$^\ell$.
From section \ref{sec:concatenated}, we know that the coding map for
this code is given by $\Omega^{(\Shor^\ell)} = \Omega^{\Shor} \circ \ldots
\circ \Omega^{\Shor} =(\Omega^{\Shor})^\ell$.  As $\Omega^{\Shor}$ takes
diagonal channels to diagonal channels , the effective channel due to
Shor$^\ell$ is also diagonal.  Let
\begin{equation}
[x_\ell(t),y_\ell(t),z_\ell(t)]
=(\Omega^\Shor)^\ell(\Ndep_t),
\end{equation}
which may be calculated using the polynomials of $\Omega^\Shor$ given in
(\ref{eqn:poly_shor}).

\begin{figure}
\includegraphics[scale=0.47]{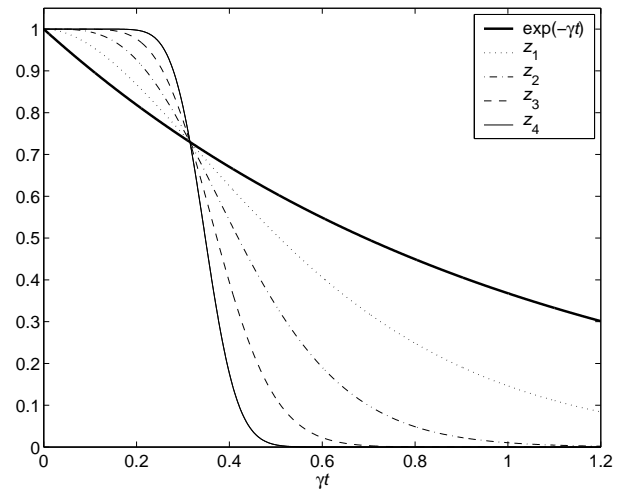}
\caption{The functions $z_\ell$, where $[x_\ell(t),y_\ell(t),z_\ell(t)]$
is the effective channel for $\ell$ concatenations of the Shor code
under the symmetric depolarizing channel.
\label{fig:shor_pointwise_Z}}
\end{figure}

The functions $z_\ell(t)$ for $0 \leq \ell \leq 4$ are plotted in
Figure  \ref{fig:shor_pointwise_Z}.  We immediately observe that as $\ell
\rightarrow \infty$ the functions $z_\ell(t)$ approach a step
function.  Denoting the step function's time of discontinuity by
$t_Z^\star$, we have $z_\ell(t) \rightarrow \theta(t_Z^\star-t)$
where
\begin{equation}
\theta(x) = \left \{ 
\begin{array}{ccc}
0 & &x < 0 \\
1 & &x > 0 
\end{array}
\right. .
\end{equation}
For $t < t_Z^\star$, each layer of concatenation decreases the
$\expval[Z]$ decoherence, yielding perfect preservation of the encoded
$\expval[Z]$ information in the infinite concatenation limit.  However,
for $t > t_Z^\star$, the $\expval[Z]$ decoherence increases.  Thus in the
infinite concatenation limit, the code will perfectly protect
$\expval[Z]$ of the logical qubit if correction is performed
prior to $t_Z^\star$; if correction is performed after this time, all
$\expval[Z]$ information is lost.

Similarly, the functions $x_\ell(t)$ and $y_\ell(t)$ approach step
function limits as $\ell \rightarrow \infty$; call the discontinuous
times of these step functions $t_X^\star$ and $t_Y^\star$.  To
perfectly protect an arbitrary state in the infinite concatenation
limit, correction must be performed prior to $t_\mathrm{th} =
\min(t_X^\star,t_Y^\star,t_Z^\star)$.  We call $t_\mathrm{th}$ the
\textit{storage threshold}.  (We use the term ``storage threshold'' to
indicate that the threshold takes into account only noise in the
register, rather than gate or measurement errors also considered in
fault-tolerant settings.)  We now show how the coding map $\Omega^\Shor$
may be used to find this threshold.

\begin{figure}
\includegraphics[scale=0.47]{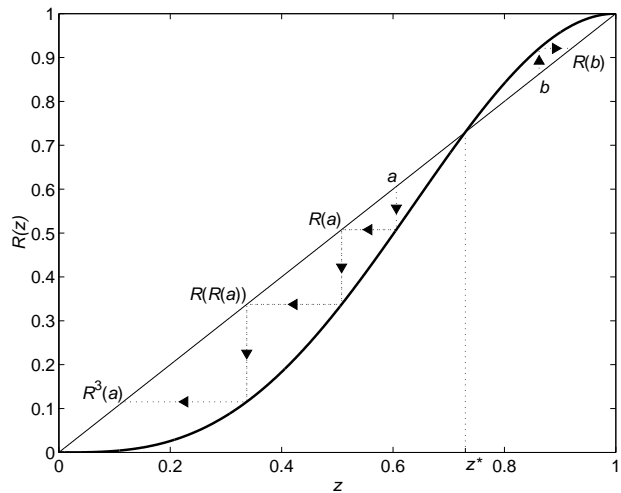}
\caption{The function $R(z)$ (plotted as the thick curve).  Observe that
the map $z \mapsto R(z)$ has fixed points at 0, 1, and $z^\star$.  The
arrows depict the iteration of this map pushing points in the interval
$(0,z^\star)$ toward 0 and points in the interval $(z^\star,1)$ toward
1.
\label{fig:R(z)}}
\end{figure}

Observe in Figure \ref{fig:shor_pointwise_Z} that the plots of $z_\ell(t)$
all intersect at a point $(\gamma t_Z^\star, z^\star)$.  Writing
$\Omega^\Shor$ in the form (\ref{eqn:form_poly_shor}), we have
$z_{\ell+1}(t) = R(z_\ell(t))$.  The function $R(z)$ is plotted in
Figure \ref{fig:R(z)}.  We see that the map $z \mapsto R(z)$ has fixed
points at $0$, $1$, and a point $z^\star \approx 0.7297$.  (We find
$z^\star$ by numerically solving $z=R(z)$ on the interval $(0,1)$.)
Iterating the map pushes points in the interval $(0,z^\star)$ toward
$0$, and pushes points in the interval $(z^\star,1)$ toward $1$.  In the
language of dynamical systems \cite{dynamical_standard}, $0$ and $1$ are
attracting fixed points, and $z^\star$ is a repelling fixed point.  This
behavior leads to the shape of the plots in
Figure \ref{fig:shor_pointwise_Z}.  We then invert $e^{-\gamma t_Z^\star}
= z^\star$ to find $\gamma t_Z^\star = 0.3151$.  The function $P(x)$ has
the same qualitative behavior on $(0,1)$ as $R(z)$, so we may similarly
find $x^\star \approx 0.9003$ and $\gamma t_X^\star \approx 0.1050$.

We cannot use the same method to find $t_Y^\star$, as $y_{\ell+1}(t)$ is
a function of $x_{\ell}(t)$, $y_{\ell}(t)$ and $z_{\ell}(t)$, not just
of $y_{\ell}(t)$ alone.  (This problem is evident from plots of the
functions $y_\ell(t)$: though these functions approach a step function
in the $\ell \rightarrow \infty$ limit, they do not all intersect at a
point as the plots of $z_\ell(t)$ do.)  However, we now argue that
finding $t_X^\star$ and $t_Z^\star$ is sufficient to find $t_Y^\star$.
For $t < t_X^\star$, $x_\ell(t)\rightarrow 1$ and $z_\ell(t) \rightarrow
1$.  Using the complete positivity constraints
(\ref{eqn:CP_constraints}), we find that if $[x,y,z]$ is a channel,
$x=z=1$ implies $y=1$.  Since the space of channels $[x,y,z]$ is
closed and bounded (it consists of the boundary and interior of a tetrahedron
in $\mathbb{R}^3$), $x_\ell(t)\rightarrow 1$ and $z_\ell(t) \rightarrow
1$ implies $y_\ell(t)\rightarrow 1$.  Now for $t_X^\star < t <
t_Z^\star$, $x_\ell(t)\rightarrow 0 $ and $z_\ell(t) \rightarrow 1$.
Using the complete positivity constraints (\ref{eqn:CP_constraints}), we
find that if $[x,y,z]$ is a channel, $x=0$ and $z=1$ implies $y=0$.
Thus we may conclude that for these values of $t$, $y_\ell(t)\rightarrow
0$.  We now have $y_\ell(t) \rightarrow 1$ for $t < t_X^\star$, and
$y_\ell(t) \rightarrow 0$ for $t > t_X^\star$, thus we conclude
$t_Y^\star = t_X^\star$.  More generally, if we know $t_X^\star$ and
$t_Z^\star$, then $t_Y^\star$ is given by $\min(t_X^\star,t_Z^\star)$.
We may therefore conclude that $\gamma t_Y^\star \approx 0.1050$, and so
$\gamma t_\mathrm{th} \approx 0.1050$.  (The value of $t_Y^\star$ could
also be obtained from the dynamics of the polynomial maps $P$, $Q$ and
$R$ without making reference to the complete positivity constraint, but
the method presented here requires less argumentation.)

We may also phrase these thresholds in the language of finitely probable
errors.  Consider the symmetric Pauli channel given by
(\ref{eqn:pauli_channel}) with $p_X=p_Y=p_Z=\frac{p}{3}$.  This channel
subjects a qubit to a random Pauli error with probability $p$, and is
described by $\Nop^\mathrm{Pauli}(p) = [1 - \frac{4}{3}p, 1 -
\frac{4}{3}p, 1 - \frac{4}{3}p]$.  Observe that the symmetric Pauli
channel and the symmetric depolarizing channel are related by
$\Nop^\mathrm{Pauli}(\frac{3}{4}(1-e^{-\gamma t})) = \Ndep_t$. Thus in the
infinite concatenation limit with $\Nop^\mathrm{Pauli}(p)$ acting on each
register qubit, the $\expval[\sigma]$ of the logical qubit will be
perfectly protected if $p < p^\star_\sigma = \frac{3}{4}(1-e^{-\gamma
t^\star_\sigma})$.  Define the threshold probability $p_\mathrm{th} =
\min\{p^\star_X,p^\star_Y,p^\star_Z\}$; for $p < p_\mathrm{th}$, all
encoded qubits are perfectly protected in the infinite concatenation
limit.  Values for $\gamma t^\star_\sigma$ and $p_\mathrm{th}$ appear in
Table \ref{tbl:asymptotics}.

\begin{table}
\begin{tabular}{|c||c|c||c|c||c||c|}
\hline
Code & \multicolumn{2}{|c||}{Shor} & \multicolumn{2}{|c||}{Shor$'$} 
     & Steane & Five-Bit\\
\hline
$\sigma$ & $X, Y$ & $Z$ & $X, Y$ & $Z$ & $X,Y,Z$ & $X,Y,Z$\\
\hline
$\gamma t_\sigma^\star$ & 0.1050 & 0.3151 & 0.1618 &  0.2150 & 0.1383 & 0.2027 \\
\hline
$p_\mathrm{th}$ & \multicolumn{2}{|c||}{0.0748} &
\multicolumn{2}{|c||}{0.1121} & 0.0969 & 0.1376
\\
\hline
\end{tabular}
\caption{Code storage threshold results.  (See text for discussion.)
\label{tbl:asymptotics}}
\end{table}

We now use similar methods to derive thresholds for the Shor$'$, Steane,
and Five-Bit codes presented in the previous section.  First, consider
the Shor$'$ code.  Let $[x'_\ell(t),y'_\ell(t),z'_\ell(t)]
=(\Omega^{\Shor'})^\ell(\Ndep_t)$.  The $y'_\ell(t)$ approach a step
function as $\ell \rightarrow \infty$, but $x'_\ell(t)$ and $z'_\ell(t)$
approach a limit cycle of period 2: we find that $x'_{2\ell}$ and
$z'_{2\ell + 1}$ both approach $\theta(t_1^\star - t)$ for some value of
$t_1^\star$, while $x'_{2\ell+1}$ and $z'_{2\ell}$ approach
$\theta(t_2^\star - t)$ for some distinct value of $t_2^\star$.  From
the form of $\Omega^{\Shor'}$ given in (\ref{eqn:form_poly_shor_prime}),
we see that $x'_{\ell+1}(t)$ is a function of $z'_{\ell}(t)$, and
$z'_{\ell+1}(t)$ a function of $x'_{\ell}(t)$, so it is not so
surprising that the sequence $z'_0,x'_1,z'_2,x'_3,\ldots$ converges and
the sequence $x'_0,z'_1,x'_2,z'_3,\ldots$ converges.  To find the
threshold, we simply consider the sequence of channels
$[x'_{2\ell},y'_{2\ell},z'_{2\ell}]$, generated by the map
$(\Omega^{\Shor'})^2$.  From (\ref{eqn:form_poly_shor_prime}) we see
that $x'_{2(\ell+1)} = R(P(x'_{2\ell}))$ and $z'_{2(\ell+1)} =
P(R(z'_{2\ell}))$.  Thus to find the values $t_X^\star$, $t_Y^\star$ and
$t_Z^\star$, we find the fixed points of the maps $x \mapsto R(P(x))$
and $z \mapsto P(R(z))$, and proceed as with the Shor code.  As shown in
Table \ref{tbl:asymptotics}, we find that, compared to the Shor code,
the Shor$'$ code has greater values for $t_X^\star$ and $t_Y^\star$, and
a lesser value for $t_Z^\star$.  As the threshold $t_\mathrm{th}$ is given
by the minimum of these three values, the Shor$'$ code outperforms the
Shor code in the infinite concatenation limit.

The map $\Omega^\Steane$, given by (\ref{eqn:poly_steane}), has the same
form as the Shor code map (\ref{eqn:form_poly_shor}), and therefore we
can use the same methods to find the Steane code thresholds.  The map
$\Omega^\Five$, given by (\ref{eqn:poly_five}), has a different form.
However, as $\Ndep_t$ has the symmetric form $[x,x,x]$ and
$\Omega^\Five$ preserves this symmetry by taking $[x,x,x]$ to
$[U(x,x,x),U(x,x,x),U(x,x,x)]$, we may find $t_X^\star = t_Y^\star =
t_Z^\star$ simply by finding the fixed point of $x \mapsto U(x,x,x)$.
Results are summarized in Table \ref{tbl:asymptotics}.  We find that the
Five-Bit code has the largest threshold, and therefore the best
performance in the infinite concatenation limit.  It is interesting to
note that the Shor$'$ code outperforms the Steane code in the infinite
concatenation limit, even though the opposite is true for only one layer
of each code.

We conclude our discussion of the thresholds by comparing the exact
values of $p_\mathrm{th}$ to those calculated with traditional
leading-order techniques (e.g. in \cite{qecc_background}).  First
consider the Five-Bit code.  Under the symmetric Pauli channel
$\Nop^\mathrm{Pauli}(p)$, a physical qubit is unmodified with probability
$1-p$.  The Five-Bit code perfectly protects its encoded information if
no more than one of the five physical qubits are subjected to a Pauli
error.  Under $\Nop^\mathrm{Pauli}(p)$, the probability of no errors on
any physical qubit is $(1-p)^5$, and the probability of exactly one
error is $5p(1-p)^4$.  We then assume that all greater-weight errors are
uncorrectable, and find that the probability of a correctable error is
$(1-p)^5 + 5p(1-p)^4 = 1 - 10p^2 + O(p^3)$.  The threshold value
$p_\mathrm{th}$ is the value of $p$ at which the single physical qubit and
the encoded information have the same probability of error.  Thus to
estimate the threshold we solve $1-10p^2 = 1-p$, yielding $p_\mathrm{th} =
\frac{1}{10}$.  Thus the leading-order calculation underestimates the
actual threshold (0.1376) by $27\%$.  (The assumption that all
errors of greater weight are uncorrectable assures that the
approximation underestimates, rather than overestimates, the threshold.)
The Steane code corrects all weight-one errors, and weight-two errors
consisting of an $X$ on one bit and a $Z$ on another bit.  A similar
calculation finds the probability of a correctable error to be $1 -
\frac{49}{3}p^2 + O(p^3)$, yielding $p_\mathrm{th} \approx 0.0612$, a
$37\%$ underestimate.  The Shor code corrects all weight-one errors, and
weight-two errors such that any $X$ and $Y$ operators occur in different
blocks, and any $Y$ and $Z$ errors occur in the same block.  The
probability of a correctable error is found to be $1 - 16p^2 + O(p^3)$,
yielding $p_\mathrm{th} = 0.0625$, a $16\%$ underestimate.  The analysis
is exactly the same for the Shor$'$ code, yet the Shor$'$ code has a
very different threshold; in this case, the leading-order result
underestimates the threshold by $44\%$.

\section{Conclusion\label{sec:conclusion}}

We have shown how a code's performance for a given error model can be
described by the effective channel for the encoded information.  The
methods presented for calculating the effective channel have allowed us
to find the performance of several codes of interest under single-bit
Pauli channels, and further have allowed us to find thresholds
describing these codes' asymptotic limits of concatenation under the
symmetric depolarizing channel.  Though we chose to restrict our
attention to diagonal channels, these methods can be applied to any
uncorrelated error model (e.g. the amplitude-damping channel
\cite{qecc_background}, which is non-diagonal), and will substantially
simplify the exact analysis of code performance in these more general
settings.

We believe that this effective channel description of code performance
may be useful in other contexts as well.  For example, this work could
be extended to take account of encoding and decoding circuit errors,
thereby providing a method for calculating exact fault-tolerant
thresholds.  Also, by providing a comprehensive method for describing
the performance of a quantum code without reference to a particular
error model (e.g. bitflip and phaseflip errors) perhaps these methods
will allow us to address open questions such as the optimal code for a
given error model, and the quantum channel capacity.

\begin{acknowledgments}
This work was partially supported by the Caltech MURI Center for Quantum
Networks and the NSF Institute for Quantum Information.  B.R. acknowledges
the support of an NSF graduate fellowship, and thanks J. Preskill
and P. Parrilo for insightful discussions.
\end{acknowledgments}

\appendix

\section{Stabilizer Codes and Diagonal Channels\label{sec:stabilizer}}

In this appendix we consider the effective channel $\Gop =
\Omega^C(\Nop^{(1)})$ when $\Nop^{(1)}$ is diagonal and $C$ is a
stabilizer code.  We show that $\Gop$ is also diagonal, and show how
the stabilizer formalism facilitates its calculation.  The reader
unfamiliar with stabilizer codes is directed to \cite{qecc_background}
for an introduction, and to \cite{Gottesman} for a more complete
discussion.

Since $\Nop^{(1)}$ is diagonal, the terms $N^{(1)}_{\nu_i \mu_i}$ in the
expression for the effective channel (\ref{eqn:G_uncorrelated}) vanish
for $\nu_i \neq \mu_i$.  Thus we have
\begin{equation}
\label{eqn:G_uncorrelated_diag}
\Gop_{\sigma \sigma'} =
\sum_{\{\mu_i\}}
\left (
\beta_{\{\mu_i\}}^{\sigma}
\alpha_{\{\mu_i\}}^{\sigma'}
\prod_{i = 1}^N \Nop^{(1)}_{{\mu_i}{\mu_i}} \right ),
\end{equation}
dramatically simplifying the calculation of $\Gop$.  The coefficients
$\alpha_{\{\mu_i\}}^{\sigma'}$ and $\beta_{\{\mu_i\}}^{\sigma}$  are
defined in terms of the $E_{\sigma'}$ and $D_\sigma$ operators in 
(\ref{eqn:E_pauli}) and (\ref{eqn:D_pauli}); to calculate these
operators we now consider the code $C$ in more detail.

Let $C$ be a stabilizer code given by stabilizer $S=\{S_k\} \subset
\pm\{I,X,Y,Z\}^{\otimes N}$, storing one qubit in an $N$-qubit register.
The stabilizer $S$ defines the codespace, and the logical operators
$\bar{I},\bar{X}, \bar{Y},\bar{Z} \subset \pm\{I,X,Y,Z\}^{\otimes N}$
determine the particular basis of codewords $\ket[\0bar]$,
$\ket[\1bar]$.  Recall that the $E_{\sigma'}$ operators act as
$\frac{1}{2}\sigma'$ on the codespace and vanish elsewhere.  It can be
shown that $P_C = \frac{1}{|S|} \sum_k S_k$ acts as a projector onto the
codespace, and by definition the logical operators $\bar{\sigma}'$ act as
$\sigma'$ on the codespace.  Thus
\begin{equation}
\label{eqn:Eops_stabilizer}
E_{\sigma'} = \tfrac{1}{2}P_C\bar{\sigma}' = \tfrac{1}{2|S|}\sum_k S_k
\bar{\sigma}'
\end{equation}
will act as $\frac{1}{2}\sigma'$ on the codespace and vanish elsewhere.

As an example, consider the bitflip code introduced in section
\ref{sec:qec_basics}.  The bitflip code may be specified as a stabilizer
code, with $S = \{III,ZZI,IZZ,ZIZ\}$, $\bar{I} = III$, $\bar{X} = XXX$,
$\bar{Y} = -YYY$, $\bar{Z} = ZZZ$.  The above expression
reproduces the expressions for the $E_{\sigma'}$ presented in
(\ref{eqn:bitflip_Eops}).  Without the stabilizer formalism, deriving
(\ref{eqn:bitflip_Eops}) is an exercise in expanding
projectors in a basis of Pauli operators; with this method the
computation is very simple.

We now construct the $D_\sigma$ operators for the stabilizer code.  As
in section \ref{sec:qec_basics}, let $\{P_j\}$ be the projectors describing
the syndrome measurement.  For a stabilizer code, the recovery operators
$R_j$ are each chosen to be a Pauli operator taking the space projected
by $P_j$ back to the codespace.  Consider the expression for $D_\sigma$
given by (\ref{eqn:Dops_nobasis}); substituting in the
expression (\ref{eqn:Eops_stabilizer}) for $E_\sigma$, we have
\begin{equation}
D_\sigma = \tfrac{1}{|S|}\sum_{k,j} R_j^\dagger S_k \bar{\sigma} R_j.
\end{equation}
Now because $R_j$, $S_k$ and $\bar{\sigma}$ are all Pauli operators,
they either commute or anti-commute.  For two Pauli operators $V$ and
$W$, let $\eta(V,W) = \pm 1$ for $VW = \pm WV$.  Commuting the $R_j$ to
the left in the above expression and noting that $R_j^\dagger R_j = 1$,
\begin{equation}
\label{eqn:Dops_stabilizer}
\begin{array}{rcl}
D_\sigma & = &
\tfrac{1}{|S|}\sum_{k,j}\eta(S_k,R_j)\eta(R_j,\bar{\sigma}) S_k
\bar{\sigma} \\
 & = & \tfrac{1}{|S|}\sum_k f_{k \sigma} S_k \bar{\sigma}
\end{array}
\end{equation}
with $f_{k \sigma} = \sum_j \eta(S_k,R_j)\eta(R_j,\bar{\sigma})$.
Again, as an example consider the stabilizer definition of the bitflip
code.  The recovery operators are $III$, $XII$, $IXI$, and $IXI$.  Evaluating
the above expression for $D_\sigma$ yields the previous result of
(\ref{eqn:bitflip_Dops}).

Using the expressions (\ref{eqn:Eops_stabilizer}) and
(\ref{eqn:Dops_stabilizer}) for the $E_{\sigma'}$ and $D_\sigma$
operators in the stabilizer formalism, we we will now find the
coefficients $\alpha_{\{\mu_i\}}^{\sigma'}$ and
$\beta_{\{\nu_i\}}^{\sigma}$ as defined in (\ref{eqn:E_pauli}) and
(\ref{eqn:D_pauli}).  Since $S_k\bar{\sigma}$ is a Pauli operator, the
sums (\ref{eqn:Eops_stabilizer}) and (\ref{eqn:Dops_stabilizer}) are
expansions of these operators $E_{\sigma'}$ and $D_\sigma$ in the Pauli
basis; if we were to write down these sums explicitly for a given
stabilizer code, the coefficients $\alpha$ and $\beta$ could be read off
immediately, e.g. from (\ref{eqn:bitflip_Eops}) and
(\ref{eqn:bitflip_Dops}).

This approach may be formalized as follows.  First, note that $S_k$ and
$\bar{\sigma}$ are both Hermitian Pauli operators, and they commute;
therefore their product is also a Hermitian Pauli operator, i.e. $S_k
\bar{\sigma} \in \pm\{I,X,Y,Z\}^{\otimes N}$.  For any operator $V = \pm
\mu_1\otimes\ldots\otimes\mu_N$ with $\mu_i \in \{I,X,Y,Z\}$, let
$\phi(V) = \mu_1\otimes\ldots\otimes\mu_N$, and let $a(V) \in \{0,1\}$
such that $V = (-1)^{a(V)}\phi(V)$.  Then, using $|S| = 2^{N-1}$, we may
re-write (\ref{eqn:Eops_stabilizer}) and (\ref{eqn:Dops_stabilizer}) as
\begin{eqnarray}
\label{eqn:Eops_stabilizer_phi}
E_{\sigma'} &=& \sum_k (-1)^{a(S_k\bar{\sigma}')}
\tfrac{1}{2^N}\phi(S_k\bar{\sigma}') \\
\label{eqn:Dops_stabilizer_phi}
D_\sigma & = & \sum_k (-1)^{a(S_k\bar{\sigma})}
\tfrac{1}{|S|}f_{k \sigma}
\phi(S_k\bar{\sigma}).
\end{eqnarray}

Comparing (\ref{eqn:Eops_stabilizer_phi}) with the definition of
$\alpha^{\sigma'}_{\{\mu_i\}}$ in (\ref{eqn:E_pauli}), we see that each
term of the sum over $k$ contributes to a single coefficient
$\alpha^{\sigma'}_{\phi(S_{k}\bar{\sigma}')}$, as
$\frac{1}{2^N}\phi(S_k\bar{\sigma}')$ is of the form $(\frac{1}{2}\mu_1
\otimes \ldots \otimes \frac{1}{2}\mu_N)$.  Similarly, each term in
(\ref{eqn:Dops_stabilizer_phi}) contributes to a single coefficient
$\beta^\sigma_{\phi(S_{k}\bar{\sigma})}$.  Lemma 2 of Appendix
\ref{sec:miscellaneous} shows that $\phi(S_k\bar{\sigma}) \neq
\phi(S_{k'}\bar{\sigma})$ unless $k=k'$ and $\sigma = \sigma'$.  Thus
each term in (\ref{eqn:Eops_stabilizer_phi}) contributes to a
\textit{distinct} coefficient
$\alpha^{\sigma'}_{\phi(S_{k}\bar{\sigma}')}$, and each term in
(\ref{eqn:Dops_stabilizer_phi}) contributes to a \textit{distinct}
coefficient $\beta^\sigma_{\phi(S_{k}\bar{\sigma})}$.  We may therefore
simply read off the coefficients from (\ref{eqn:Eops_stabilizer_phi})
and (\ref{eqn:Dops_stabilizer_phi}), yielding
\begin{eqnarray}
\label{eqn:alpha_stabilizer}
\alpha^{\sigma'}_{\phi(S_k \bar{\sigma}')} & = & (-1)^{a(S_k\bar{\sigma}')} \\
\label{eqn:beta_stabilizer}
\beta^{\sigma}_{\phi(S_k \bar{\sigma})} &=&
(-1)^{a(S_k\bar{\sigma})}\tfrac{1}{|S|}f_{k\sigma},
\end{eqnarray}
and all other $\alpha^{\sigma'}_{\{\mu_i\}}$ and 
$\beta^{\sigma}_{\{\mu_i\}}$ vanishing.

We now evaluate $\Gop = \Omega^C(\Nop^{(1)})$ where $\Nop^{(1)}=[x,y,z]$
using (\ref{eqn:G_uncorrelated_diag}).  The only non-vanishing terms
$\beta_{\{\mu_i\}}^{\sigma}$ occur when $\mu_1\otimes\ldots\otimes\mu_N
= \phi(S_k \bar{\sigma})$ for some $k$ and $\sigma$, and the only the
only non-vanishing terms $\alpha_{\{\mu_i\}}^{\sigma'}$ occur when
$\mu_1\otimes\ldots\otimes\mu_N = \phi(S_k \bar{\sigma}')$ for some $k$
and $\sigma'$.  Thus the coefficients $\beta_{\{\mu_i\}}^{\sigma}
\alpha_{\{\mu_i\}}^{\sigma'}$ of (\ref{eqn:G_uncorrelated_diag}) will
vanish unless $\mu_1\otimes\ldots\otimes\mu_N = \phi(S_k \bar{\sigma}) =
\phi(S_{k'} \bar{\sigma}') $ for some $k$ and $k'$.  As proved in
Lemma 2 of Appendix \ref{sec:miscellaneous}, this cannot happen
when $\sigma \neq \sigma'$.  Thus all the matrix elements $\Gop_{\sigma
\sigma'}$ vanish when $\sigma \neq \sigma'$, i.e. $\Gop$ is diagonal.

Having demonstrated that the coding map $\Omega^C$ of a stabilizer code
$C$ takes diagonal channels to diagonal channels, and because $\Gop_{II}
= 1$ from trace preservation, we need only compute $\Gop_{XX}$,
$\Gop_{YY}$, and $\Gop_{ZZ}$ using (\ref{eqn:G_uncorrelated_diag}) to
find $\Gop = \Omega^C([x,y,z])$.  These computations can be performed
using the methods of section \ref{sec:coding_map}, but we conclude
this section by expressing these elements using the stabilizer
formalism, which may be computationally advantageous.

Consider the diagonal terms $\Gop_{\sigma \sigma}$ given by
(\ref{eqn:G_uncorrelated_diag}).  We need only sum over the
non-vanishing coefficients $\alpha$ and $\beta$, which are given by
(\ref{eqn:alpha_stabilizer}) and (\ref{eqn:beta_stabilizer}).  Substituting
in these expressions yields 
\begin{equation}
\Gop_{\sigma \sigma} =
\sum_{k}
\left (
\frac{1}{|S|}f_{k \sigma}
\prod_{i = 1}^N \Nop^{(1)}_{\phi_i(S_k \bar{\sigma})
\phi_i(S_k \bar{\sigma})}\right )
\end{equation}
where $\phi_i(V)$ denotes $\mu_i$ for $\phi(V) = \mu_1 \otimes \ldots
\otimes \mu_N$.  Now as $\Nop^{(1)} = [x,y,z]$, the product of the matrix
elements of $\Nop^{(1)}$ in the previous expression is simply a product of
$x$'s, $y$'s and $z$'s; each factor appears as many times as
(respectively) $X$, $Y$ and $Z$ appear in $\phi(S_k \bar{\sigma})$.
Letting $w_\sigma(p)$ denote the $\sigma$-weight of a Pauli operator
$p$, e.g. $w_X(XYX) = 2$, we have
\begin{equation}
\Gop_{\sigma\sigma'}=
\delta_{\sigma\sigma'}
\frac{1}{|S|}\sum_k f_{k\sigma}
x^{w_X(S_k\bar{\sigma})} y^{w_Y(S_k\bar{\sigma})}
z^{w_Z(S_k\bar{\sigma})}.
\end{equation}

\section{}
\label{sec:miscellaneous}

This appendix contains lemmas deferred from previous sections.

\textbf{Lemma 1:} The decoding operation $\Dop$ given by
(\ref{eqn:decoding_kraus}) is a quantum operation.

\textit{Proof:} 
From (\ref{eqn:decoding_kraus}) we have
\begin{equation}
\Dop[\rho] = \sum_j B^\dagger A_j \rho A_j^\dagger B.
\end{equation}
To prove that $\Dop$ is a quantum operation, we must show
\begin{equation}
\sum_{j}(A_j^\dagger B)(B^\dagger A_j) = \mathbf{1}
\end{equation}
where $\mathbf{1}$ is the identity on the register space.
As we assumed that $\mathcal{R}$ maps all states into the
codespace, we can choose the operators $A_j$ to only have range on the
codespace.  With such a choice, $A_j^\dagger A_j = A_j^\dagger
(\ketbra[\0bar,\0bar]+\ketbra[\1bar,\1bar])A_j = A_j^\dagger BB^\dagger
A_j$.  We therefore have
\begin{equation}
\sum_j (A_j^\dagger B)(B^\dagger A_j) = \sum_j A_j^\dagger A_j.
\end{equation}
From (\ref{eqn:recovery_kraus}) we have $\sum_j A_j^\dagger A_j =
\mathbf{1}$, and so $\Dop$ is a quantum operation.  {\scriptsize
$\blacksquare$}

\textbf{Lemma 2:} For a stabilizer code with stabilizer $\{S_k\}$ and
logical operators $\{\bar{\sigma}\}$, and $\phi$ defined in Appendix
\ref{sec:stabilizer}, $\phi(S_k\bar{\sigma}) \neq
\phi(S_{k'}\bar{\sigma}')$ unless $k=k'$ and $\sigma = \sigma'$.  

\textit{Proof:} Suppose we have $\phi(S_k\bar{\sigma}) =
\phi(S_{k'}\bar{\sigma}')$; then $S_k\bar{\sigma} =
\pm S_{k'}\bar{\sigma}'$.  As the stabilizers $S_k$ and $S_{k'}$
act trivially on the codespace, $S_k\bar{\sigma}$ and $\pm
S_{k'}\bar{\sigma}'$ act respectively as $\sigma$ and $\pm\sigma'$ on
the codespace.  Thus we must have $\sigma = \pm\sigma'$, which requires
$\sigma = \sigma'$ and the $\pm$ sign be positive.  We now have
$S_k\bar{\sigma} = S_{k'}\bar{\sigma}$; right-multiplying by
$\bar{\sigma}$ yields $S_k = S_{k'}$, and thus $k = k'$.  {\scriptsize
$\blacksquare$}


\begin{thebibliography}{14}
\expandafter\ifx\csname natexlab\endcsname\relax\def\natexlab#1{#1}\fi
\expandafter\ifx\csname bibnamefont\endcsname\relax
  \def\bibnamefont#1{#1}\fi
\expandafter\ifx\csname bibfnamefont\endcsname\relax
  \def\bibfnamefont#1{#1}\fi
\expandafter\ifx\csname citenamefont\endcsname\relax
  \def\citenamefont#1{#1}\fi
\expandafter\ifx\csname url\endcsname\relax
  \def\url#1{\texttt{#1}}\fi
\expandafter\ifx\csname urlprefix\endcsname\relax\def\urlprefix{URL }\fi
\providecommand{\bibinfo}[2]{#2}
\providecommand{\eprint}[2][]{\url{#2}}

\bibitem[{\citenamefont{Shor}(1995)}]{qecc_original}
\bibinfo{author}{\bibfnamefont{P.~W.} \bibnamefont{Shor}},
  \bibinfo{title}{``Scheme for Reducing Decoherence in Quantum Computer Memory''},
  \bibinfo{journal}{Phys. Rev. A} \textbf{\bibinfo{volume}{52}},
  \bibinfo{pages}{2493} (\bibinfo{year}{1995}).

\bibitem[{\citenamefont{Steane}(1996)}]{steane}
\bibinfo{author}{\bibfnamefont{A.~M.}~\bibnamefont{Steane}},
\bibinfo{title}{``Error Correcting Codes in Quantum Theory''},
  \bibinfo{journal}{Phys. Rev. Lett.} \textbf{\bibinfo{volume}{77}},
  \bibinfo{pages}{793} (\bibinfo{year}{1996}).

\bibitem[{\citenamefont{Nielsen and Chuang}(2000)}]{qecc_background}
\bibinfo{author}{\bibfnamefont{M.~A.} \bibnamefont{Nielsen}} \bibnamefont{and}
  \bibinfo{author}{\bibfnamefont{I.~L.} \bibnamefont{Chuang}},
  \emph{\bibinfo{title}{Quantum Computation and Quantum Information}}
  (\bibinfo{publisher}{Cambridge University Press}, \bibinfo{year}{2000}),
  \bibinfo{note}{and references therein; J. Preskill, Lecture Notes,
  http://theory.caltech.edu/$\sim$preskill/ph219 (1998)}.

\bibitem[{\citenamefont{Preskill}(1997)}]{ftc_background}
\bibinfo{author}{\bibfnamefont{J.}~\bibnamefont{Preskill}},
  \bibinfo{title}{``Fault-tolerant Quantum Computation''},
  \eprint{quant-ph/9712048}
  (\bibinfo{year}{1997}), \bibinfo{note}{and references therein}.

\bibitem[{\citenamefont{Knill and Laflamme}(1996)}]{concatenation_original}
\bibinfo{author}{\bibfnamefont{E.}~\bibnamefont{Knill}} \bibnamefont{and}
  \bibinfo{author}{\bibfnamefont{R.}~\bibnamefont{Laflamme}},
  \bibinfo{title}{``Concatenated Quantum Codes''},
  \eprint{quant-ph/9698012} (\bibinfo{year}{1996}). 

\bibitem[{\citenamefont{Knill and Laflamme}(1997)}]{theory_qecc}
\bibinfo{author}{\bibfnamefont{E.}~\bibnamefont{Knill}} \bibnamefont{and}
  \bibinfo{author}{\bibfnamefont{R.}~\bibnamefont{Laflamme}},
  \bibinfo{title}{``Theory of Quantum Error-Correcting Codes''},
  \bibinfo{journal}{Phys. Rev. A} \textbf{\bibinfo{volume}{55}},
  \bibinfo{pages}{900} (\bibinfo{year}{1997}).

\bibitem[{\citenamefont{Kraus}(1983)}]{kraus}
\bibinfo{author}{\bibfnamefont{K.}~\bibnamefont{Kraus}},
  \emph{\bibinfo{title}{States, Effect, and Operations}}
  (\bibinfo{publisher}{Springer-Verlag}, \bibinfo{year}{1983}).

\bibitem[{\citenamefont{Ruskai et~al.}(2001)\citenamefont{Ruskai, Szarek
  et~al.}}]{Ruskai}
\bibinfo{author}{\bibfnamefont{M.~B.} \bibnamefont{Ruskai}},
  \bibinfo{author}{\bibfnamefont{S.}~\bibnamefont{Szarek}},
  \bibnamefont{et~al.}, 
  \bibinfo{title}{``An Analysis of Completely-Positive Trace-Preserving
Maps on $\mathcal{M}_2$''}, \eprint{quant-ph/0101003} (\bibinfo{year}{2001}).

\bibitem[{\citenamefont{Lidar et~al.}(2001)\citenamefont{Lidar, Bacon
  et~al.}}]{collective}
\bibinfo{author}{\bibfnamefont{D.~A.} \bibnamefont{Lidar}},
  \bibinfo{author}{\bibfnamefont{D.}~\bibnamefont{Bacon}},
  \bibnamefont{et~al.}, 
  \bibinfo{title}{``Decoherence-Free Subspaces for Multiple-Qubit Errors I: Characterization''},
\bibinfo{journal}{Phys. Rev. A}
  \textbf{\bibinfo{volume}{63}}, \bibinfo{pages}{022306}
  (\bibinfo{year}{2001}).

\bibitem[{\citenamefont{Devaney}(1989)}]{dynamical_standard}
\bibinfo{author}{\bibfnamefont{R.~L.} \bibnamefont{Devaney}},
  \emph{\bibinfo{title}{An Introduction to Chaotic Dynamical Systems}}
  (\bibinfo{publisher}{Addison-Wesley}, \bibinfo{year}{1989}).

\bibitem[{\citenamefont{King and Ruskai}(2001)}]{King}
\bibinfo{author}{\bibfnamefont{C.}~\bibnamefont{King}} \bibnamefont{and}
\bibinfo{author}{\bibfnamefont{M.~B.} \bibnamefont{Ruskai}},
\bibinfo{title}{``Minimal Entropy of States Emerging from Noisy Quantum
Channels''}, \bibinfo{journal}{IEEE Trans. Info. Theory}
\textbf{\bibinfo{volume}{47}}, \bibinfo{pages}{192}
(\bibinfo{year}{2001}), \eprint{quant-ph/9911079}.

\bibitem[{\citenamefont{Gottesman}(1997)}]{Gottesman}
\bibinfo{author}{\bibfnamefont{D.}~\bibnamefont{Gottesman}},
\bibinfo{title}{``Stabilizer Codes and Quantum Error Correction''},
  \bibinfo{note}{Ph.D. thesis, Caltech} 
  (\bibinfo{year}{1997}), \eprint{quant-ph/9705052}.

\bibitem[{\citenamefont{Laflamme et~al.}(1996)\citenamefont{Laflamme, Miqel
  et~al.}}]{laflamme_miqel}
\bibinfo{author}{\bibfnamefont{R.}~\bibnamefont{Laflamme}},
  \bibinfo{author}{\bibfnamefont{C.}~\bibnamefont{Miqel}},
  \bibnamefont{et~al.}, 
  \bibinfo{title}{``Perfect Quantum Error Correcting Code''},
\bibinfo{journal}{Phys. Rev. Lett.}
  \textbf{\bibinfo{volume}{77}}, \bibinfo{pages}{198} (\bibinfo{year}{1996}).

\bibitem[{\citenamefont{Bennett et~al.}(1996)\citenamefont{Bennett, DiVincenzo
  et~al.}}]{bennett_divincenzo}
\bibinfo{author}{\bibfnamefont{C.~H.} \bibnamefont{Bennett}},
  \bibinfo{author}{\bibfnamefont{D.~P.} \bibnamefont{DiVincenzo}},
  \bibnamefont{et~al.}, 
  \bibinfo{title}{``Mixed-State Entanglement and Quantum Error Correction''},
\bibinfo{journal}{Phys. Rev. A}
  \textbf{\bibinfo{volume}{54}}, \bibinfo{pages}{3824} (\bibinfo{year}{1996}).

\end{thebibliography}
\end{document}